\documentclass[10pt,journal,compsoc]{IEEEtran}

\usepackage[utf8]{inputenc}   
\usepackage[T1]{fontenc}
\usepackage[english]{babel}

\usepackage{amsmath,amsfonts}
\usepackage{algorithmic}
\usepackage{algorithm}
\usepackage{array}
\usepackage[caption=false,font=normalsize,labelfont=sf,textfont=sf]{subfig}
\usepackage{textcomp}
\usepackage{stfloats}
\usepackage{verbatim}
\usepackage{graphicx}
\usepackage{xspace}
\usepackage[skins,breakable]{tcolorbox}
\usepackage{booktabs} 
\usepackage{multirow} 
\usepackage{rotating}
\usepackage[hyphens]{url} 
\usepackage[breaklinks=true]{hyperref}
\usepackage{siunitx}
\usepackage{tabularx}
\usepackage[inline]{enumitem}
\usepackage[numbers, sort]{natbib}
\usepackage{makecell}
\usepackage{marvosym}
\usepackage[flushleft]{threeparttable}
\usepackage{listings}
\usepackage{balance}

\usepackage{upquote}

\newcounter{policy}
\newcommand{\newpolicy}{\refstepcounter{policy}P\thepolicy}
\newcounter{example}
\newcommand{\newexample}{\refstepcounter{example}E\theexample}

\newtcolorbox[auto counter]{examplebox}[1][]{
colback=gray!5,        
  colframe=black!30,     
  boxrule=0.3pt,         
  fonttitle=\itshape,    
  title={Examp. \thetcbcounter},
  #1
}

\newtcolorbox{summarybox}[1]{
  enhanced,
  colback=white,         
  colframe=black,        
  boxrule=0.8pt,         
  fonttitle=\bfseries,   
  fontupper=\small,
  title={Summary #1:}
}

\definecolor{grayrow}{rgb}{0.7,0.7,0.7}

\usepackage{csquotes}
\newcommand*{\enq}[1]{\enquote{{\itshape#1}}} 

\lstnewenvironment{plain} {\lstset{
  numbers=none
}}{}

\begin{document}

\title{Self-Admitted GenAI Usage in Open-Source~Software}
\author{Tao Xiao, Youmei Fan, Fabio Calefato, Christoph Treude,\\ Raula Gaikovina Kula, Hideaki Hata, Sebastian Baltes~\Letter
\thanks{Sebastian Baltes is the corresponding author.}

\IEEEcompsocitemizethanks{\IEEEcompsocthanksitem{T. Xiao is with Kyushu University, Japan.
E-mail: xiao@ait.kyushu-u.ac.jp}
\IEEEcompsocthanksitem{Y. Fan is with Nara Institute of Science and Technology, Japan.
E-mail: fan.youmei.fs2@is.naist.jp}
\IEEEcompsocthanksitem{F. Calefato is with University of Bari, Italy.
E-mail: fabio.calefato@uniba.it}
\IEEEcompsocthanksitem{C. Treude is with Singapore Management University, Singapore.
E-mail: ctreude@smu.edu.sg}
\IEEEcompsocthanksitem{R. G. Kula is with University of Osaka, Japan.
E-mail: raula-k@ist.osaka-u.ac.jp}
\IEEEcompsocthanksitem{H. Hata is with Shinshu University, Japan.
E-mail: hata@shinshu-u.ac.jp}
\IEEEcompsocthanksitem{S. Baltes is with Heidelberg University, Germany.
E-mail: sebastian.baltes@uni-heidelberg.de}}}

\maketitle

\begin{abstract}
The widespread adoption of generative AI (GenAI) tools such as GitHub Copilot and ChatGPT is transforming software development.
Since generated source code is virtually impossible to distinguish from manually written code, their real-world usage and impact on open-source software (OSS) development remain poorly understood.
In this paper, we introduce the concept of \emph{self-admitted GenAI usage}, that is, developers explicitly referring to the use of GenAI tools for content creation in software artifacts.
Using this concept as a lens to study how GenAI tools are integrated into OSS projects, we analyze a curated sample of more than 200,000 GitHub repositories, identifying 1,292 such self-admissions across 156 repositories in commit messages, code comments, and project documentation.
Using a mixed methods approach, we derive a taxonomy of 32 tasks, 10 content types, and 11 purposes associated with GenAI usage based on 1,292 qualitatively coded mentions.
We then analyze 13 documents with policies and usage guidelines for GenAI tools and conduct a developer survey to uncover the ethical, legal, and practical concerns behind them.
Our findings reveal that developers actively manage how GenAI is used in their projects, highlighting the need for project-level transparency, attribution, and quality control practices in AI-assisted software development.
Finally, we examine the longitudinal impact of GenAI adoption on \emph{code churn} in 151 repositories with self-admitted GenAI usage and find no general increase, contradicting popular narratives on the impact of GenAI on software development.
\end{abstract}


\section{Introduction}
\label{sec:introduction}

\IEEEPARstart{T}he emergence of generative artificial intelligence (GenAI) tools such as ChatGPT and GitHub Copilot has redefined software development, as documented by literature reviews~\cite{hou2023large}, developer studies~\cite{Vaithilingam2022, Liang2024}, and productivity studies~\cite{DBLP:journals/cacm/ZieglerKLRRSSA24}.
These tools assist developers in writing and reviewing code, refining documentation, and automating various aspects of the software development lifecycle.
Although prior research has evaluated the technical capabilities of GenAI tools~\cite{Nguyen2022} and surveyed their usability~\cite{Liang2024}, only a few studies have systematically investigated their real-world adoption and usage patterns, for example, by mining GenAI mentions in repositories~\cite{Tufano2024}, analyzing AI-generated pull request descriptions~\cite{xiao2024generative}, or compiling datasets of developer-ChatGPT conversations~\cite{10.1145/3643991.3648400}.
One reason is that only the tool vendors have access to fine-grained usage data~\cite{DBLP:journals/cacm/ZieglerKLRRSSA24} that allows them to determine which code suggestions were accepted and hence which code was co-authored by GenAI tools.
Without additional context, generated code is virtually impossible to distinguish from human-authored code.

As a result, much of what we know about GenAI usage in software development is inferred indirectly, either from vendor-controlled telemetry, laboratory studies, or analyses of repository activity that rely on aggregated metrics rather than direct evidence of GenAI use. This makes it difficult to understand how GenAI tools are actually integrated into collaborative development workflows, how their use is governed in practice, and how their impact should be interpreted at the project level. In real-world settings, developers must decide how much to rely on or revise AI-generated content, and maintainers must determine whether to prohibit, restrict, or encourage GenAI use.
Researchers increasingly rely on aggregate metrics such as code churn to assess claims about software quality degradation.
Without observable, project-level signals of GenAI usage embedded in software artifacts, these decisions risk being shaped by assumptions, anecdotal evidence, or broad industry narratives rather than by empirical data. Studying explicit references to GenAI usage offers a way to ground these discussions in real development practices, making it possible to examine not only what GenAI is used for, but also how it is acknowledged, regulated, and followed by human action in open-source software (OSS) projects.

These projects, with their collaborative nature and publicly accessible repositories~\cite{dabbish2012social}, offer a unique context to study the adoption of GenAI tools.
Although such tools promise to support OSS projects by automating development tasks, there are also reports of ``AI slop'' wasting valuable time of maintainers~\cite{curl-ai-slop} or GenAI-generated contributions leading to more rework~\cite{gitclear2024}.

We introduce the concept of \textbf{self-admitted GenAI usage}, inspired by the notion of self-admitted technical debt~\cite{potdar2014exploratory}. 
Just as developers acknowledge technical debt through comments and commits, they sometimes explicitly refer to using GenAI tools. These self-admissions can highlight tasks delegated to GenAI tools, challenges encountered, or changes made due to AI-generated content. 
Identifying such usage enabled us to explore three research questions (RQs).
First, to understand the practical applications of GenAI tools in software development, we examined which \emph{tasks} (e.g., writing test cases) are supported or automated by these tools, which \emph{contents} (e.g., methods in source files) are referenced in GenAI-related mentions, and which \emph{purposes} (e.g., acknowledging GenAI use) such mentions serve. We then investigated the \emph{regulation and recommendation} of GenAI usage, and concluded with an analysis of \emph{code churn} following the first GenAI mention in a project.

\begin{description}[style=multiline, leftmargin=8mm]
\item[RQ1] \emph{For which tasks, contents, and purposes do open-source developers mention GenAI tools?}
\end{description}

One finding that emerged was that project maintainers have begun to establish policies and usage guidelines regarding their use (see Table~\ref{tab:reg}).
These regulations provide insights into emerging best practices, ethical considerations, and potential concerns surrounding GenAI adoption.
Understanding project-level policies is crucial for the responsible integration of GenAI tools in collaborative software development, leading to our second RQ:

\begin{description}[style=multiline, leftmargin=8mm]
\item[RQ2] \emph{How do open-source projects regulate or recommend the usage of GenAI tools?}
\end{description}

In addition to understanding how developers use GenAI tools and how projects regulate their usage, it is important to understand their impact on software quality and maintenance.
The 2024 GitClear report~\cite{gitclear2024}, which received considerable attention in the developer community, claimed that increased code churn after GenAI adoption indicates \enq{downward pressure on code quality.}
The report defines code churn as \enq{the percentage of lines that are reverted or updated less than two weeks after being authored,} interpreting such changes as \enq{either incomplete or erroneous when the author initially wrote, committed, and pushed them} to the repository. 
To investigate this claim, we formulate a third RQ:

\begin{description}[style=multiline, leftmargin=8mm]
\item[RQ3] \emph{Does the code churn change after open-source projects start using GenAI tools?}
\end{description}

We conducted a large-scale empirical study of more than 200,000 OSS repositories hosted on GitHub.
Our investigation focused on identifying explicit mentions of GenAI tools in various artifacts and analyzing how these mentions relate to development activities.
We followed a mixed methods approach, combining a qualitative analysis of GenAI-related mentions with a quantitative examination of code churn over time, resulting in four main contributions:

\begin{enumerate}
    \item We introduce self-admitted GenAI usage as an empirical lens for studying GenAI adoption in open-source software and curate a dataset of 1,292 self-admitted GenAI usages across 156 GitHub repositories.
    \item Using a mixed-methods approach, we derive a taxonomy of GenAI usage comprising 32 development tasks, 10 content types, and 11 purposes, grounded in a qualitative analysis of the identified usages.
    \item We empirically analyze how open-source projects govern GenAI usage by examining 13 policies and usage guidelines, contextualized through a developer survey.
    \item We assess the longitudinal impact of GenAI adoption on software evolution using a repository-level analysis of code churn in 151 projects, showing that GenAI adoption does not lead to a general increase in churn and that effects are stronger for generation tasks.
\end{enumerate}

\section{Methodology}
\label{sec:methodology}

\begin{figure*}
    \centering
    \includegraphics[width=\linewidth]{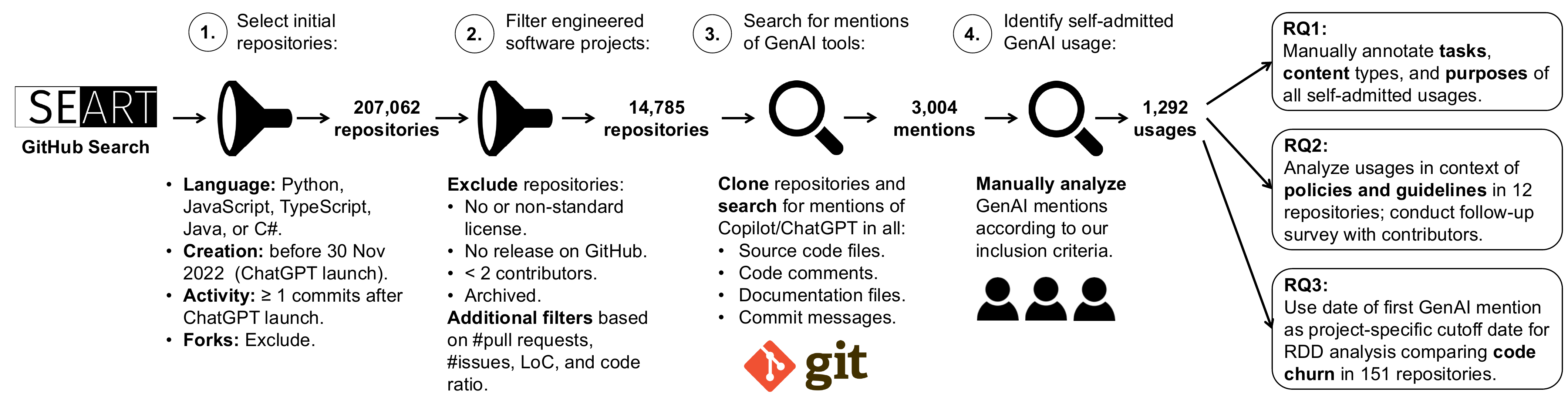}
    \caption{Overview of the data collection process used to answer our three research questions, from the selection (1) and filtering (2) of GitHub repositories to the extraction of GenAI mentions (3) and the identification of self-admitted GenAI usage (4) in these repositories.}
    \label{fig:over}
\end{figure*}

We followed a mixed-methods research design.
Our data collection process is visualized in Figure~\ref{fig:over}.
After retrieving instances of self-admitted GenAI usage from open-source GitHub repositories, we conducted a qualitative analysis to answer \textbf{RQ1}. Through multiple iterative coding phases, we labeled these instances to classify supported tasks and generated content.
Since this qualitative analysis yielded a considerable number of statements that focused on the regulation or recommendation of GenAI practices, we followed up with a closer analysis of these aspects as part of \textbf{RQ2}.
For \textbf{RQ3}, we used self-admitted GenAI usages to approximate the time when the projects started using GenAI tools, to analyze the effect of GenAI usage on code churn using a Regression Discontinuity Design (RDD).

\subsection{Repository Sampling}

The foundation of our research is a large sample of open-source GitHub repositories.
We selected GitHub as our study platform because it is the largest and most widely used open-source hosting service, with over 500 million repositories according to the 2024 Octoverse report~\cite{ghreport}, making it the most suitable environment for analyzing trends of GenAI usage in open-source software development.
Using the \href{https://seart-ghs.si.usi.ch/}{GitHub search tool} provided by Dabic et al.~\cite{Dabic:msr2021data}, we selected repositories primarily written in the five most popular programming languages as of the above-mentioned report~\cite{ghreport}: Python, JavaScript, TypeScript, Java, and C\#.
This focus on the most popular languages ensures that our study is both manageable and relevant to the most commonly used development ecosystems.
Since \textbf{RQ3} aims at a comparison of code churn before and after projects started using GenAI tools, we only selected repositories that: (1) were created before the ChatGPT launch date (30 November 2022) and (2) had at least one commit on or after this date.
Moreover, to eliminate duplicates, we excluded forks. 
Our initial sample of GitHub projects contained 207,062 repositories distributed across Python (77,542), JavaScript (48,500), TypeScript (37,424), Java (25,160), and C\# (18,436).

Since our interest is to study ``engineered'' software projects~\cite{DBLP:journals/ese/MunaiahKCN17}, we applied three additional filtering criteria.
First, we excluded repositories not declaring a license or using non-standard licenses (marked as \emph{Other} in the GitHub search tool).
For the remaining repositories, we labeled all 38 distinct licenses we found and then removed projects declaring licenses not commonly used for software projects.
These licenses included \emph{Creative Commons Attribution 4.0 International}, \emph{Creative Commons Zero v1.0 Universal}, \emph{Creative Commons Attribution Share Alike 4.0 International}, and the \emph{SIL Open Font License 1.1}.
Second, we excluded repositories without any release on GitHub, fewer than two contributors, and those marked as \emph{archived}.
Third, we filtered the repositories based on an analysis of various descriptive statistics.
We analyzed the distribution of central repository properties per programming language.
The properties we considered were the number of pull requests, the number of issues, and the repository size measured in lines of code (as provided by the GitHub search tool).

To select engineered software projects with sufficient development data, we excluded repositories in the first quartile ($Q_1$) for each metric, therefore removing the lowest 25\%. Furthermore, we excluded repositories with a code ratio (defined as $lines\_of\_code / (lines\_of\_code + lines\_of\_comments)$) outside the 97\% confidence interval. The rationale behind this threshold is that engineered software projects are usually documented using source code comments. Filtering out repositories beyond the 97\% confidence interval helps eliminate outliers, that is, repositories with very little code, or codebases dominated by code without comments. 
A sanity check further confirmed that this ratio serves as a reliable indicator for filtering out non-software or poorly structured projects.

Our final sample of GitHub repositories, obtained in February 2024, contained 14,785 GitHub repositories distributed across Java (5,060), C\# (3,544), TypeScript (2,464), Python (1,875), and JavaScript (1,842). Table~\ref{tab:stat} provides descriptive statistics for the studied GitHub repositories.

\begin{table}[tb]
\caption{Descriptive statistics for studied GitHub repositories ($n=14,785$).}
\label{tab:stat}
\centering
\resizebox{\columnwidth}{!}{
\begin{tabular}{lrrrrr}
\toprule
                 & \textbf{Min} & \textbf{Mean}      & \textbf{Median} & \textbf{Max} & \textbf{Std Dev.} \\
\midrule
\textbf{\# Issues}        & 9       & 372    & 98     & 171,039 & 1,880  \\
\textbf{\# Pull Requests} & 9       & 454    & 134    & 38,421 & 1,334   \\
\textbf{\# Contributors}  & 2       & 30     & 15     & 475   & 49   \\
\textbf{\# Lines of Code} & 1,800    & 135,223 & 25,399  & 47,165,225 & 734,845\\
\textbf{Code Ratio (\%)} & 4.300     & 78.274     & 78.721  & 99.996 & 12  \\
\bottomrule
\end{tabular}}
\end{table}

\subsection{Identifying Self-Admitted GenAI Usages}
\label{sec:identifying-mentions}

\begin{table}
\caption{File extensions we included when searching for mentions of GenAI tools in our sample of GitHub repositories.}
\label{tab:file-extensions}
\centering
\begin{tabularx}{\linewidth}{llX}
\toprule
\textbf{Type} & \textbf{Language} & \textbf{File Extensions} \\
\midrule
Code & Python & \texttt{.py}, \texttt{.ipynb} \\
Code & Java & \texttt{.java}, \texttt{.jsp} \\
Code & TypeScript & \texttt{.ts}, \texttt{.tsx}, \texttt{.vue} \\
Code & JavaScript & \texttt{.js}, \texttt{.jsx}, \texttt{.vue}, \texttt{.mjs}, \texttt{.cjs} \\
Code & C\# & \texttt{.cs}, \texttt{.aspx}, \texttt{.cshtml} \\
Doc. & All & \texttt{.md}, \texttt{.markdown}, \texttt{.mdown}, \texttt{.mkdn}, \texttt{.mkd}, \texttt{.mdwn}, \texttt{.mdtxt}, \texttt{.mdtext}, \texttt{.txt}, \texttt{.text}, \texttt{.adoc}, \texttt{.asciidoc}, \texttt{.rst}, \texttt{.textile}, \texttt{.dbk}\\
\bottomrule
\end{tabularx}
\end{table}

To identify self-admitted GenAI usage in our filtered sample of GitHub repositories, we retrieved mentions of the two most popular GenAI tools among developers as of the 2023 Stack Overflow Developer Survey \cite{StackOve71:online}. In that survey, ChatGPT was identified as the most popular general AI tool and GitHub Copilot as the most popular AI developer tool.
Then, in the second step, we annotated these mentions to identify those related to content generation.
We wrote a Python script for the following process:

{\small
\begin{enumerate}
    \item \textbf{Clone} the default branch of the repository.
    \item Search all \textbf{source code files} for mentions of ChatGPT or Copilot within \textbf{code comments}; save the complete comments along with their language (i.e., the natural language such as English or Chinese).
    \item Search all \textbf{documentation files} for mentions of ChatGPT or Copilot; save the lines in which the mentions were found, again along with their language.
    \item Search all \textbf{commit messages} for mentions of ChatGPT or Copilot; save the corresponding commit messages along with their language.
\end{enumerate}}

An initial analysis of all files in the repositories revealed a large number of false positive matches, that is, mentions of GenAI tools that were not related to content generation.
Therefore, we decided to focus on specific file types when searching for mentions in source code and documentation files.
We derived these lists based on common file extensions for the particular programming languages, as well as an analysis of all unique file extensions in which we found mentions during our first data collection run (see Table~\ref{tab:file-extensions}).
We further decided to only search mentions of GenAI tools in source code comments, not across the whole source code.
This is because, during our initial analysis, we found many false positives that were not related to content generation but to code that calls APIs related to ChatGPT or Copilot.
We developed heuristics to reduce these false positives, which we outline in the following.

For identifying mentions of GenAI tools, we employed regular expressions with the following pattern:
{\small
\begin{verbatim}
re.compile(r'(.?)' + llm_tool + r'(.?)',
  re.IGNORECASE | re.DOTALL) 
\end{verbatim}
}

\noindent where the variable \verb|llm_tool| was assigned the value \verb|r'chat[ \-_]{0,1}gpt'| for ChatGPT and \verb|r'co[ \-_]{0,1}pilot'| for Copilot. 
These patterns allowed us to capture variations in how these tools were referenced while minimizing false positives.
We developed heuristics to further reduce the number of false positives. For example, we noticed that in false positive matches, the mentions of GenAI tools were often surrounded by commas or underscores, e.g., when they were part of URLs for API calls.
Our supplementary material contains the full source code that documents our retrieval approach.

Running the above retrieval process on all repositories yielded 3,004 mentions of GenAI tools: 1,572 in commit messages, 397 in source code comments, and 1,035 in documentation files.
These mentions were automatically obtained using regular expressions and filtered according to heuristics.
However, they still included mentions that were not related to content generation. 
Thus, we conducted a thorough manual inspection of all mentions to eliminate false positives.
This review process was guided by the following instructions: 

{\small
\begin{enumerate}
    \item We \textbf{include} mentions indicating that content was generated using ChatGPT or Copilot and then copied into the repository. We use a broad definition of ``content'' that includes not only source code but also comments, translations, and other textual elements.
    \item For commits, we also \textbf{include} mentions that indicate a modification of previously generated content (e.g., a refactoring or fix for previously generated content) or commits that remove comments indicating the usage of ChatGPT or Copilot to generate content.
    \item For documentation files, we \textbf{include} mentions that indicate content generation, discuss or regulate the usage of ChatGPT or Copilot in the repositories, and mentions that acknowledge the use of these tools. 
\end{enumerate}}

To evaluate the coding instructions, two authors independently labeled a sample of mentions, deciding whether they should be included or not.
We calculated a sample size of 341 mentions (of 3,004) to achieve estimates with a 95\% confidence level and a 5\% confidence interval.
The inspection resulted in disagreement between the two authors for only 14 cases (4\% of the sample).
The two authors discussed these cases and tried to reach a consensus. During these discussions, a third author helped resolve each disagreement and suggested possible improvements to the categories.
To assess inter-rater reliability, we computed Fleiss' kappa~\cite{fleiss1971measuring} by applying bootstrap resampling methods with 1,000 iterations. The resulting 95\% confidence interval was estimated to be (0.87, 0.95), indicating an \enq{almost perfect} agreement.
Given this high agreement, the first author continued to inspect the remaining mentions alone.
In total, we identified 1,292 true-positive mentions of GenAI~tools that were aligned with our inclusion criteria. We found true-positive mentions in 156 repositories (11 Python, 12 JavaScript, 37 TypeScript, 47 C\#, and 49 Java repositories). We did not merge mentions referring to the same GenAI action (e.g., a commit message and a code comment referring to the same change), as they may indicate distinct usage patterns.

\subsection{Data and Code Availability}

To enable replication and future research, we have prepared supplementary material that includes the filters we used to sample GitHub repositories, the raw data we retrieved, the manually labeled GenAI tool mentions, the Python scripts we used for data retrieval and analysis, and the questionnaires used for our developer survey.
The package is available online~\cite{xiao_2025_15871468}.

\section{Reasons for Mentioning GenAI Tools}

To answer \textbf{RQ1}, we qualitatively analyzed the GenAI mentions that we collected and curated, categorizing them according to tasks, contents, and purposes.

\subsection{Method}
\label{sec:RQ1A}

We performed an open-coding methodology combined with card sorting to manually analyze our sample of 1,292 GenAI tool mentions (see Section~\ref{sec:identifying-mentions}).
The initial coding~\cite{charmaz2014constructing} involved systematically examining and categorizing the data according to emerging conceptual themes.
In our study, this involved analyzing individual GenAI tool mentions to identify recurring patterns and assign corresponding codes.
Following this initial coding phase, we performed open card sorting to organize low-level codes into higher-level abstract categories, allowing us to recognize broader themes and relationships (focused coding).
Three authors of this paper collaborated throughout this process to ensure a rigorous and consistent annotation.

A preliminary analysis identified 1,008 mentions related to Copilot for Pull Requests.\footnote{\url{https://githubnext.com/projects/copilot-for-pull-requests}}
Of these, 1,000 instances occurred in a single repository (\texttt{pancakeswap/pancake-frontend}) where commit messages directly reused pull request descriptions generated by the tool, while the remaining eight mentions in other repositories explicitly documented that developers used Copilot for Pull Requests to generate pull request descriptions, without reusing those descriptions as commit messages. In addition, we identified one separate case in which contributors were encouraged to use ChatGPT to produce pull request descriptions (P\ref{p:8}); this case was not part of the Copilot for Pull Requests group.

Given the overrepresentation of a single, highly repetitive use case, namely the reuse of Copilot-generated pull request descriptions as commit messages, we set aside these 1,008 mentions during the initial round of coding to avoid skewing the development of the coding schema. After establishing a stable set of categories based on the remaining data, we revisited these deferred cases and incorporated them into the analysis.

To build the code book, two authors independently analyzed 284 GenAI mentions. 
The code book development was guided by the following questions:

{\small
\begin{itemize}
    \item \textbf{Task:} \textit{Which task has the GenAI tool supported or automated?} Tasks include, for example, writing a test case, fixing a bug, and refactoring the code base.
    \item \textbf{Content:} \textit{Which content is the GenAI mention referring to?} Content categories include methods in source files, sections in documentation files, and commit messages.
    \item \textbf{Purpose:} \textit{Why has the GenAI tool been mentioned?} Possible purposes include acknowledgment of usage for code generation and regulation of usage within the project.
\end{itemize}}

Our coding process allowed coders to assign multiple codes per mention.
During the iterative refinement of the codes and categories, we observed an interesting pattern in how developers describe their work with GenAI tools.
Each mention typically encompasses two distinct but interconnected perspectives: (i) the specific task delegated to the GenAI tool and (ii) the broader development task the human developer aims to accomplish.
To capture this pattern, we split the task-related codes into two sub-categories: \textbf{GenAI task} and \textbf{developer task}.
We provide the final code book and code assignment as part of our supplementary material.

Using Fleiss' kappa~\cite{fleiss1971measuring}, we assessed the interrater reliability between the two coders. 
The analysis yielded ``substantial'' to ``almost perfect'' agreement levels on task ($k = 0.81-0.89$), content ($k = 0.95-0.99$), and purpose ($k = 0.79-0.92$), according to standard guidelines for interpreting $k$~\cite{viera2005understanding}.
Through iterative discussions, the two coders worked to achieve consensus on the categorizations, with a third researcher arbitrating unresolved disagreements and recommending refinements to the categories.
The first author then independently checked the 1,008 mentions that we had initially deferred.

\subsection{Results}
\label{sec:RQ1R}

\begin{table*}[tb]
\caption{GenAI-assisted tasks (\textbf{RQ1}): Definition and frequency of categories and codes (n = 284 + 1,000 + 8 = 1,292); the code \emph{PR description} is counted and discussed separately because most of it only occurred in one repository (see Section~\ref{sec:tasks}).}
\label{tab:task}
\centering
\begin{tabularx}{\textwidth}{lp{2.5cm}Xr}
\toprule
\textbf{Category} & \textbf{Code} & \textbf{Definition} & \textbf{\#}\\
\midrule

\multirow{19}{*}{\textbf{Generation}} 
 & Code & Understand coding tasks written in natural language and generate corresponding code. & 105 \\
 \cline{2-4}
 & Test data & Create test input/output based on software requirements or the existing codebase. 
 & 9 \\
 \cline{2-4}
 & Comment & Generate code comments that explain the purpose and logic of code blocks. & 9 \\
 \cline{2-4}
 & Test file & Create test files based on software requirements or the existing codebase. & 8 \\
\cline{2-4}
 & Regex & Craft regular expressions tailored to specific text matching needs. & 6 \\
 \cline{2-4}
 & README & Create README files that provide essential information, e.g., project descriptions. & 4 \\
 \cline{2-4}
 & Dummy text & Produce placeholder text that mimics real content in style, structure, and format. & 4 \\
 \cline{2-4}
 & Test method & Create test methods based on software requirements or the existing codebase. 
 & 2 \\
 \cline{2-4}
 & Code review & Generate reviews that suggest improvements and identify potential issues in code changes. & 2 \\
 \cline{2-4}
 & {Commit message} & Generate commit messages that summarize code changes. & 2 \\
 \cline{2-4}
 & Tutorial & Produce instructional content on specific topics and step-by-step guidance for projects. & 2 \\
 \cline{2-4}
 & Zod schema & Create Zod schemas in TypeScript and JavaScript for type safety and data validation. & 2 \\
 \cline{2-4}
 & Test class & Create test classes based on software requirements or the existing codebase. 
 & 1 \\
 \cline{2-4}
 & {Coding practices} & Generate guidelines and best practices for coding in the projects. & 1 \\
 \cline{2-4}
 & Variable & Suggest meaningful variable names that improve code semantics and readability. 
 & 1 \\
 \cline{2-4}
 & Changelog & Compile changelogs that document changes, features, and fixes in new software versions. & 1 \\
 \cline{2-4}
 & Configuration & Generate project-specific configuration files, e.g., performance and security settings. & 1 \\
 \cline{2-4}
  & Text & Generate general text that is not mentioned above. & 8 \\
 \cline{2-4}
& \emph{PR description}  & \emph{Create explicit PR descriptions to assist understanding the changes in the PRs.} & \emph{1,009} \\

  \midrule

 \multirow{3}{*}{\textbf{Translation}} & Text & Convert text between different languages, e.g., software internationalization (i18n). & 49 \\
 \cline{2-4}
 & Code & Convert code between different programming languages, preserving the original logic and functionality while adapting to the syntax and idiomatic patterns of the target language. & 1 \\
 \midrule
 
 \multirow{2}{*}{\textbf{Optimization}} & {Code refactoring} & Restructure code without altering its functionality, aiming to make the code maintainable. & 29 \\
 \cline{2-4}
 & {Code improvement} & Improve existing code, mention is accompanied by ``improve.'' & 5 \\
 \midrule
  
\multirow{10}{*}{\textbf{Maintenance}} & {Label revision} & Analyze, update, and improve text labels, ensuring clarity, accuracy, and consistency. 
& 8 \\
\cline{2-4}
 & {README revision} & Analyze, update, and improve README files, ensuring clarity, accuracy, and consistency. 
 & 7 \\
  \cline{2-4}
 & {Document revision} & Analyze, update, and improve documents, ensuring clarity, accuracy, and consistency. & 4 \\
 \cline{2-4}
 & {Changelog revision} & Analyze, update, and improve changelogs, ensuring clarity, accuracy, and consistency.
 & 2 \\
  \cline{2-4}
 & {Prompt refinement} & Optimize and clarify the prompts to elicit the most relevant and accurate responses. & 1 \\
 \cline{2-4}
 & {Color suggestion} & Suggest color schemes for UI/UX design based on best practices and design requirements. & 1 \\
  \cline{2-4}
 & {Dependency\newline upgrade} & Analyze software dependencies and suggest updates to ensure compatibility and security while minimizing breaking changes. & 1 \\
 \cline{2-4}
 & {Version update} & Suggest meaningful version numbering for software releases for systematic version control. & 1 \\
  \cline{2-4}
 & {Comment revision} & Analyze, update, and improve code comments, ensuring clarity, accuracy, and consistency. 
 & 1 \\
 \midrule
  
\textbf{Other} & - & Operate general functionality, like Q\&A and blog generation. & 12 \\
\midrule
 
\textbf{None} & - & There is no specific task for the GenAI tool. & 9 \\
\bottomrule
\end{tabularx}
\end{table*}

Our analysis of mentions revealed distinct patterns in how developers integrate GenAI tools into their development workflows. 
In the following, we describe the categories and codes capturing development tasks, content types, and usage purposes, which emerged from our analysis.

\subsubsection{GenAI-Assisted Tasks}
\label{sec:tasks}

\begin{table}[tb]
\centering
\caption{Examples of self-admitted GenAI usage referenced in this paper.}
\label{tab:examples}
\begin{tabular}{lll}
\toprule
ID & Artifact & Link \\
\midrule
\newexample\label{e:1} & commit & \href{https://github.com/aksio-insurtech/cratis/commit/e97eee5163653bd6f3f2feb1b0c24955285c8f26}{aksio-insurtech/cratis/commit/e97e...} \\
\newexample\label{e:2} & comment & \href{https://github.com/iportalteam/immersiveportalsmod/blob/1.20.4/src/main/java/qouteall/imm_ptl/core/portal/shape/RectangularPortalShape.java\#L95}{iportalteam/imm.../PortalShape.java\#L95} \\
\newexample\label{e:3} & commit & \href{https://github.com/fusion-flux/portal-cubed/commit/0a9d6deafead0e16ac58ef9ac1e554dc8a6edd95}{fusion-flux/portal-cubed/commit/0a9d...} \\
\newexample\label{e:4} & commit & \href{https://github.com/vercel/next.js/commit/d21025cc3a50e2ff8a7137d5d5c94576218f01e7}{vercel/next.js/commit/d210...} \\
\newexample\label{e:5} & commit & \href{https://github.com/pancakeswap/pancake-frontend/commit/4e0f034a0129e9800b572fa5fda4453130733d07}{pancakeswap/pancake.../commit/4e0f...} \\
\newexample\label{e:6} & doc. & \href{https://github.com/pancakeswap/pancake-frontend/blob/develop/CONTRIBUTING.md}{pancakeswap/.../CONTRIBUTING.md} \\
\newexample\label{e:7} & comment & \href{https://github.com/LAMP-Platform/LAMP/blob/22f20cf12f608bb237fa5eaa22ee9971e9d09eee/YAM2E/Utilities/Format.cs\#L171}{LAMP-Platform/LAMP/.../Format.cs\#L171}\\
\newexample\label{e:8} & doc. & \href{https://github.com/ant-design/ant-design/blob/fa3fddb0edd38251524e9b4606c74f013f91f500/docs/blog/github-actions-workflow.en-US.md?plain=1\#L101}{ant-des.../github-actions-workflow.en-US.md} \\
\newexample\label{e:9} & doc. & \href{https://github.com/Minecraft-AMS/Carpet-AMS-Addition/blob/b52cf767a9c0efc9392f86c17a9d680ac7a68266/README\_en.md?plain=1\#L38}{Minecraft-AMS/Carpet-.../README\_en.md} \\
\newexample\label{e:10} & comment & \href{https://github.com/BdR76/CSVLint/blob/65b8c46fcaf357bf17b99b9e921c95f341ac7a02/CSVLintNppPlugin/CsvLint/CsvGenerateCode.cs\#L733-L735}{BdR76/.../CsvGenerateCode.cs\#L733-L735} \\
\newexample\label{e:11} & commit & \href{https://github.com/VelvetToroyashi/Silk/commit/35d9bf9dafada4dc89d7e6c1c3617be7b93aefe4}{VelvetToroyashi/Silk/commit/35d9...} \\
\newexample\label{e:12} & commit & \href{https://github.com/deephaven/web-client-ui/commit/d852e495a81c26a9273d6f8a72d4ea9fe9a04668}{deephaven/web-client-ui/commit/d852...} \\
\newexample\label{e:13} & comment & \href{https://github.com/hypar-io/Elements/blob/5ec3391069aa9d02d9f3a1f4fca9eebe5bbc6260/Elements/src/Geometry/Ellipse.cs\#L166-L167}{hypar-io/elements/.../Ellipse.cs\#L166-L167} \\
\newexample\label{e:14} & comment & \href{https://github.com/DominoKit/domino-ui/blob/ebe51ac3117676d24fad5f67f6941b3b81687d5b/domino-ui/src/main/java/org/dominokit/domino/ui/sliders/Slider.java\#L546-L550}{dominokit/domino-.../Slider.java\#L546-L550}\\
\newexample\label{e:15} & doc. & \href{https://github.com/Anime4000/IFME/blob/326fe6d8c0333826a02b18d1c44b32fe9d678205/changelog.txt\#L210}{Anime4000/IFME/.../changelog.txt\#L210}\\
\newexample\label{e:16} & commit & \href{https://github.com/dotnet/project-system/commit/3aa25a5bae9309daf813302cfc2e3dddd19ea842}{dotnet/project-system/commit/3aa2...}\\
\newexample\label{e:17} & commit & \href{https://github.com/ediwang/moonglade/commit/a185a00fa9577a88ac7caeec6708ff7677c4e28f}{ediwang/moonglade/commit/a185...} \\
\bottomrule
\end{tabular}
\end{table}

Our analysis identified 32 distinct task categories in which developers use GenAI tools in their workflows. 
Table~\ref{tab:task} presents these categories along with their definitions and usage frequencies, while Table~\ref{tab:examples} shows a list of examples of self-admitted GenAI usage. 
Unsurprisingly, excluding PR-related activities, generation tasks were most common, with code generation being particularly prominent (105 instances).
Translation followed with 50 instances, while optimization and maintenance tasks accounted for 34 and 26 instances, respectively.

As mentioned above, we distinguish between developer tasks and GenAI tasks.
While Table~\ref{tab:task} lists the GenAI tasks, we also want to discuss human tasks related to GenAI tasks.
For example, in one commit message that we analyzed (E\ref{e:1}) the developer acknowledged that the code was written \enq{a bit hasty on previous release} due to \enq{trust in GitHub Copilot.}
The developer task described in the commit message was \emph{bug fixing}, while the initial task that the GenAI tool supported was \emph{code generation}.

We identified 20 mentions exhibiting this pattern of human actions triggered by an earlier GenAI action.
Among them, 13 referred to code that was initially generated using GenAI tools and then changed.
The most common follow-up activity was to fix bugs in AI-generated code (9).
In other cases, changes were reverted (1), AI-generated comments were deleted (2), or the generated code was commented out (1).
For example, one developer commented out code generated by Copilot with the note: \enq{Note: do not trust GitHub Copilot. It may use z as up axis} (E\ref{e:2}).
Another developer reverted a commit that was created with the help of ChatGPT: \enq{Revert `ChatGPT' This reverts commit 71e3...} (E\ref{e:3}).

In addition to the 13 human actions that followed AI code generation that we discussed above, we found seven human actions following the generation of configuration and validation files or an unclear role of the GenAI tool.
In five cases, developers specified restrictions or exclusions regarding GenAI usage without mentioning a specific task.
In two other cases, they removed and rewrote AI-generated configurations or validations.
For instance, one pull request superseded another that \enq{heavily relies on GitHub Copilot (which makes the progress slow and tedious)} (E\ref{e:4}).
The developer manually replaced the generated validation schema with a handwritten version.

Recent research has shown that using AI-generated PR descriptions reduces review time and increases PR merge rates~\cite{xiao2024generative}.
We found that developers reused generated PR descriptions as part of their commit messages. As mentioned in Section~\ref{sec:RQ1A}, this approach was very common in one particular project, which contributed 1,000 such mentions to our sample. These generated messages are not limited to this one project---similar patterns appear in popular projects such as 
\texttt{pytorch/pytorch} and \texttt{hasura/graphql-engine}. 
They are added when developers use the PR description as the message for merge or squash commits. This practice represents a form of explicit, self-admitted GenAI usage, embedding a clear marker of AI contribution directly into the software project's official history.
Note that, although this single use case contributes a large absolute count, it represents only one entry in our taxonomy of GenAI tasks (Table~\ref{tab:task}) and does not affect our broader findings. To illustrate this particular use case, we include an excerpt below (E\ref{e:5}).
Interestingly, the linked contribution guidelines (E\ref{e:6}) do not discuss GenAI usage.

\small{\begin{plain}
chore: Remove no used deps (#7349)
<!--
Before opening a pull request, please read the
[contributing guidelines](https://github.com/...]
first
-->
<!--
copilot:all
-->
### <samp>Generated by Copilot at b3683ce</samp>
[...]
\end{plain}}

Of the 1,009 instances of code \emph{PR description} in Table~\ref{tab:task}, 1,000 originated from a single repository and are discussed separately above. The remaining nine cases include eight instances in which developers explicitly documented the use of Copilot for Pull Requests to generate PR descriptions (Section~\ref{sec:RQ1A}), and one instance suggesting that contributors use ChatGPT to produce PR descriptions (P\ref{p:8}).

Below, we discuss the most prevalent supported tasks besides generating PR descriptions.
As expected, code generation was one of the most common GenAI-supported tasks that we observed.
Some self-admitted GenAI usage for code generation was straightforward, such as the following statement that we found in the source code comment documenting a method written in C\#: \enq{This function was written with Chat-GPT} (E\ref{e:7}).

Beyond code generation, developers used GenAI tools to generate other software artifacts, including test data or documentation.
Besides generation, GenAI tools were also used to automate code review, for example, as part of GitHub Actions workflows (E\ref{e:8}): \enq{Recently, the team has added ChatGPT to GitHub Actions to perform GenAI-based code review. The specific job can be found in the chatgpt-cr.yml file.}

After generation, translation emerged as the second most prevalent task in our analysis.
Most mentions referred to translation between natural languages. One mention referred to translation between programming languages.
An important use case was internationalization, helping developers overcome language barriers (E\ref{e:9}): \enq{Due to my limited proficiency in English, all English document translations are currently provided by ChatGPT, including this sentence.}
The one mention related code translation documented the translation of existing Python code to R (E\ref{e:10}): \enq{The following R code was generated using ChatGPT based on the Python code.}
However, the developer at the same time asked others to support them in improving the code: \enq{If anyone can refactor it to something more readable or more sensible code, please let me know or submit as a pull request.}

Code optimization represented the third largest category. Developers not only acknowledged GenAI tools usage but sometimes even thanked the tools in their commit messages (E\ref{e:11}): \enq{Forgot tabs. Thanks, Copilot.}
In addition to code, GenAI tools were also used to improve UI elements (E\ref{e:12}): \enq{I asked chatGPT to help me brainstorm improvements to some of the labels and hint text based on the [...] Interface Guidelines. I then edited them as human to improve them further.}
Interestingly, also in this case, the developer asked other members to review the generated content: \enq{Review and let me know if you think any are worse or weird.}
This, together with the human corrective actions triggered by GenAI actions we observed, points to the importance of human oversight in GenAI-assisted development.

\subsubsection{Content Types} 

\begin{table*}[tb]
\caption{Content types (\textbf{RQ1}): Definition and frequency of categories and codes.}
\centering
\scriptsize
\label{tab:content}
\begin{tabularx}{\textwidth}{lp{43mm}Xr}
\toprule
\textbf{Category} & \textbf{Code} & \textbf{Definition} & \textbf{\#}  \\
\midrule
\textbf{Project metadata} & Commit messages & Target commit messages. & 1,003 \\
\midrule
\multirow{6}{*}{\textbf{Source files}} & Whole methods & Target source code files, ranging from entire functions within a file. & 47 \\
\cline{2-4}
 & Blocks within one source code file & Target source code files, spanning multiple blocks within a single source code file. & 45 \\
 \cline{2-4}
 & One block within one source code file & Target source code files, spanning one block within a single source code file. & 39 \\
 \cline{2-4}
 & Blocks within multiple source code files & Target source code files, spanning multiple source code files. & 21 \\
 \cline{2-4}
 & Whole files & Target source code files, ranging across the entire file. & 12 \\
 \cline{2-4}
 & Whole classes & Target source code files, ranging across the entire class in the file. & 12 \\
\midrule
\multirow{3}{*}{\textbf{Project assets}} & Documentation files & Target documentation files, which include technical documents in software projects. & 106 \\
\cline{2-4}
 & Configuration files & Target configuration files, which define the operational parameters and settings. & 24 \\
 \cline{2-4}
 & Resource files & Target resource files, e.g., images, localization strings, and other binary data. & 5 \\
 \bottomrule
\end{tabularx}
\end{table*}

\begin{table*}[tb]
\caption{Purposes of GenAI usage (\textbf{RQ1}): Definition and frequency of categories and codes.}
\label{tab:purpose}
\centering
\scriptsize
\begin{tabularx}{\textwidth}{lp{34mm}Xr}
\toprule
\textbf{Category} & \textbf{Code} & \textbf{Definition} & \textbf{\#} \\
\midrule
 \multirow{4}{*}{\makecell{\textbf{Documentation and}\\\textbf{Acknowledgment}}} & Acknowledgement of usage & Recognizing and documenting the use of GenAI tools within the codebase. & 1,236 \\
 \cline{2-4}
 & Acknowledge that the bug fix is related to AI-generated code & Noting in the documentation or comments that a particular bug fix pertains to issues originating from AI-generated code. & 13 \\
  \cline{2-4}
 & Removal of Copilot comment & Deletion of comments initially suggested by GenAI tools that are no longer relevant or correct. & 2 \\
\midrule
 \multirow{5}{*}{\makecell{\textbf{Guidance and} \\\textbf{Best Practices}}} & Set example & Providing usage examples to illustrate how GenAI tools can be used. & 25 \\
 \cline{2-4}
 & Exclusion of usage within the project & Documenting rules or guidelines on how GenAI tools should not be used within the project to maintain consistency and quality. & 18 \\
 \cline{2-4}
 & Regulation of usage within the project & Documenting rules or guidelines on how GenAI tools should be used within the project to maintain consistency and quality. & 10 \\
\midrule
 \multirow{4}{*}{\textbf{Quality Assurance}} & Look for refactoring/reviewing/improving & Marking sections of content generated by GenAI tools that need to be refactored, reviewed, or improved for better performance, readability, or maintainability. & 11 \\
 \cline{2-4}
 & Warning & Issuing cautions in the code, e.g., vulnerabilities, deprecated methods, or unstable features. & 10 \\
 \cline{2-4}
 & TODO & Indicating LLM tasks that need to be completed in the future. & 2 \\
\midrule
 \multirow{2}{*}{\textbf{GenAI~Limitations}} & Blame Copilot & Specifically attributing errors or suboptimal code to suggestions made by a GenAI tool. & 3 \\
 \cline{2-4}
 & Revert & Noting the need to undo LLM changes that have led to issues or did not perform as expected. & 1 \\
\bottomrule
\end{tabularx}
\end{table*}

Our analysis identified three main categories of AI-generated content in open-source software projects, organizing ten distinct codes (see Table~\ref{tab:content}). 
Although, as mentioned before, commit messages related to Copilot PR activities dominated our dataset with over 1,000 mentions from a single repository, examining the remaining data revealed important patterns. 
Developers frequently use GenAI tools to modify source files (176 mentions).
However, other file types, such as documentation and configuration files, were also targeted (135 mentions). 

When working with source files, developers usually focus on smaller elements such as individual functions or code blocks instead of complete files.
For example, we found blocks of code implementing geometrical transformation, for which the developers added a comment indicating ChatGPT usage.
Interestingly, they even documented the prompt in the source code comment (E\ref{e:13}): \enq{Code generated from chatgpt with the following prompt:[...].}
In another example, a developer added an interface for UI elements, mentioning ChatGPT as the author in the comment (E\ref{e:14}): \enq{A functional interface to handle slider slide events. [...] @author ChatGPT.}

For project assets other than source code, GenAI was also used to generate changelogs (E\ref{e:15}): \enq{Note: This changelog is improved by OpenAI ChatGPT from my broken English input.}
Another use case we observed was adding comments explaining options in a configuration file (E\ref{e:16}): \enq{These strings were provided by GitHub Copilot. I checked the first few, and they were correct.}

\subsubsection{Purposes of GenAI Usage}

Our analysis identified 11 different purposes for GenAI mentions in software projects, grouped into four main categories (see Table~\ref{tab:purpose}).
Documentation and acknowledgment of GenAI usage emerged as the most frequent purpose.
This manifested itself in several ways, such as offering guidance (53 mentions), flagging areas needing attention (23 mentions), and addressing GenAI limitations (4 mentions).

Self-admission of GenAI usage, as illustrated by the previously mentioned comment for the generated C\# method (E\ref{e:13}), appeared consistently across projects.
Besides generation, code refactoring is another use case for mentioning GenAI usage: \enq{code refact by github copilot} (E\ref{e:17}).

Quality assurance emerged as another key purpose, with developers often requesting peer review of AI-generated content.
More examples of this can be found in Section~\ref{sec:tasks}.

\begin{summarybox}{RQ1}
For the 1,292 GenAI mentions we analyzed, developers mainly used GenAI tools for code generation, natural language translation, and code refactoring.
Source code and documentation files were the dominant generation targets.
Acknowledgment of GenAI usage was a common purpose, sometimes combined with warnings about possible negative implications.
Another important purpose was regulation (see \textbf{RQ2}).
Our analysis revealed patterns of corrective actions following code generation.
Our findings show that GenAI tools are actively used in open-source software and that developers are working on guiding their usage.
\end{summarybox}

\section{Existing Guidelines for GenAI Usage}
\label{sec:rq2}

One topic that emerged while answering \textbf{RQ1} is that some open-source projects have specific policies and guidelines around GenAI usage.
Therefore, as part of \textbf{RQ2}, we investigated how projects prohibit, restrict, or support the usage of GenAI tools.
In addition to analyzing the policies and guidelines, we conducted a survey with open-source developers to understand their views on GenAI regulation. 

\subsection{Method}

Using our sample of GenAI mentions, we found 28 mentions related to policies and usage guidelines around GenAI tool usage.
We grouped them into three groups: (1) prohibitive, (2) restrictive, and (3) supportive usage.
Table~\ref{tab:reg} presents detailed examples drawn from 13 documentation files and commit messages in 12 GitHub repositories, where the last column indicates the number of mentions identified in the software artifact.

First, we closely examined these policies and usage guidelines to understand how exactly projects regulate GenAI usage.
Second, we conducted a developer survey that included excerpts from the policies and guidelines we found.
The primary goals of the survey were to: (1) collect developer perceptions on the need for GenAI tool guidance (e.g., documenting prompts or annotating generated content) and understand the actions taken on this content before integration or publication; and (2) investigate the rationale behind policies and usage guidelines.
To investigate the second part, we asked participants if they contributed to one of the repositories from which we extracted policies and guidelines (see Table~\ref{tab:reg}) and then showed the corresponding guidelines, asking them to elaborate on the rationale behind them.
In this way, we received feedback on P\ref{p:1} and P\ref{p:10}. For developers who did not identify as contributors to any of the repositories, we showed P\ref{p:3}, P\ref{p:4}, and P\ref{p:7}, deliberately selected as examples of prohibitive and restrictive usage, and asked for their feedback on these policies. Supportive policies were not included because they are less likely to raise questions about regulation or project governance. To keep the survey focused and concise, we prioritized policies that reflect the tensions and risks surrounding GenAI usage in OSS projects. Future work could complement this by investigating developer perceptions of supportive policies and the conditions under which projects actively promote GenAI use.

Our target population was the contributors of the 12 GitHub repositories in our dataset that contained explicit GenAI usage policies (see Table~\ref{tab:reg}).
Of these, seven had the GitHub Discussions feature enabled, which we used as our primary outreach channel to gather direct developer feedback.
For the remaining five repositories where this feature was not available, as well as two repositories where our discussion posts received no response, we identified contributors on GitHub and then, to comply with GitHub's terms of service, looked for contact details outside of GitHub (e.g., personal websites or social media profiles). 
We were able to determine the email addresses of 30 contributors.
In total, we received eight survey responses, which we analyzed using a combination of open coding and card sorting.
Informed consent was obtained.
The survey questionnaire is available in our supplementary material.

\subsection{Results}

\begin{table*}[tb]
\caption{Policies and guidelines for GenAI usage in software projects (\textbf{RQ2}): \underline{\textbf{pro}}hibitive, \underline{\textbf{res}}trictive, and \underline{\textbf{sup}}portive policies (P); the last column (\#) shows the number of mentions in the corresponding documentation file or commit message.}
\label{tab:reg}
\centering
\scriptsize
\begin{tabularx}{\linewidth}{llp{15mm}Xr}
\toprule
\textbf{Goal} & \textbf{ID} & \textbf{Repository} & \textbf{Excerpt} & \# \\
\midrule
\multirow{6}{*}[-8em]{\underline{\textbf{pro}}}
& \newpolicy\label{p:1} & jqwik-team/\newline jqwik &
  \enq{jqwik Contributor Agreement - You \textbf{have authored 100\% of the contents of your contribution}. Among other things that means that you \textbf{have not used GitHub Copilot or a similar LLM} to create all or parts of your contribution! The reason is that the copyright consequences of training an LLM with mostly public code repositories have not been clarified.} (\href{https://github.com/jqwik-team/jqwik/blob/30aa7a637c460e481e842b12a961e9966d150012/CONTRIBUTING.md?plain=1\#L5-L7}{CONTRIBUTING.md}) & 1\\ \cline{2-5}
& \newpolicy\label{p:2} & jqwik-team/\newline jqwik &
    \enq{Including GH Copilot clause in CONTRIBUTING.md} (\href{https://github.com/jqwik-team/jqwik/commit/6cdc49504e526f7bef34fea9d416dc5daa8eaf33}{commit/6cdc...}) & 1\\ \cline{2-5}
& \newpolicy\label{p:3} & shoelace-style/\newline shoelace &
  \enq{\textbf{AI-generated Code} As an open-source maintainer, I respectfully ask that you \textbf{refrain from using AI-generated code} when contributing to this project. This includes code generated by tools such as GitHub Copilot, even if you make alterations to it afterwards. While some of Copilot's features are indeed convenient, \textbf{the ethics surrounding which codebases the AI has been trained on and their corresponding software licenses remain very questionable and have yet to be tested in a legal context}. I realize that one cannot reasonably enforce this any more than one can enforce not copying licensed code from other codebases, nor do I wish to expend energy policing contributors. I would, however, like to \textbf{avoid all ethical and legal challenges that result from using AI-generated code}. As such, I respectfully ask that you refrain from using such tools when contributing to this project. At this time, I will \textbf{not knowingly accept any code that has been generated in such a manner}.} (\href{https://github.com/shoelace-style/shoelace/blob/fb59fda70ed737c92611051b49bc7e3a5fed5dc5/docs/pages/resources/contributing.md?plain=1\#L30-L32}{contributing.md}) & 1\\ \cline{2-5}
& \newpolicy\label{p:4} & turms-im/turms &
  \enq{\textbf{Can Responses Generated by a Model Similar to ChatGPT be Used for Discussion?} \textbf{ChatGPT is an excellent memorizer, but its analysis of various technical solutions is quite naive}. Engaging in discussions with ChatGPT responses only \textbf{reflects a lack of critical thinking and a lack of responsibility towards the projects}. Therefore, whether we should answer such responses depends on the proportion of responses after removing ChatGPT answers. [...] \textbf{How to Identify Responses Generated by a Model Similar to ChatGPT} [...]} (\href{https://github.com/turms-im/turms/blob/0002c493ef47a0e0cd15a3de09c2cc936f710a8d/turms-docs/src/community/index.md?plain=1\#L56-L70}{index.md}) & 9\\ \cline{2-5}
& \newpolicy\label{p:5} & katsutedev/\newline mal4j &
  \enq{PLEASE READ BEFORE SUBMITTING PR \textbf{Does not include AI generated code}, such as GitHub Copilot or ChatGPT.} (\href{https://github.com/KatsuteDev/Mal4J/blob/35af4bb0ebaddc79397c7147ba46cb9ba58433b4/.github/pull_request_template.md?plain=1\#L12}{pull\_request\_template.md}) & 1\\ \cline{2-5}
& \newpolicy\label{p:6} & shred/acme4j &
  \enq{Acceptance Criteria These criteria must be met for a successful pull request: … You confirm that you \textbf{did not use AI based code generators} like GitHub Copilot for \textbf{your contribution}.} (\href{https://github.com/shred/acme4j/blob/ec726f6859b12ed59830e6e80a50daf5f034345c/CONTRIBUTING.md?plain=1\#L5-L15}{CONTRIBUTING.md}) & 1\\
\midrule
\multirow{7}{*}[-3em]{\underline{\textbf{res}}}
& \newpolicy\label{p:7} & graycoreio/\newline daffodil &
  \enq{Submitting a Pull Request (PR) Before you submit your Pull Request (PR) consider the following guidelines: Please note: If your PR contains \textbf{code that was generated by an AI tool such as ChatGPT or Copilot}, \textbf{you must disclose this} in the description of your PR.} (\href{https://github.com/graycoreio/daffodil/blob/c30b3081c8f61a0048cb6c34a8ad2256e3fdcb9e/CONTRIBUTING.md?plain=1\#L111}{CONTRIBUTING.md}) & 1\\ \cline{2-5}
& \newpolicy\label{p:10} & owasp/\newline wrongsecrets &
  \enq{\textbf{Why you should be careful with AI (or ML) and secrets} Any AI/ML solution that relies on \textbf{your input} might use that input for further improvement. This is sometimes referred to as \enq{Reinforcement learning from human feedback} … This means that when you use those and give them feedback or agree on sending them data to be more effective in helping you, then this data resides with them and might be queryable by others.} (\href{https://github.com/OWASP/wrongsecrets/blob/fb4ed66d796e6dc50e2158f3d4adea37f142fef1/src/main/resources/explanations/challenge32_reason.adoc?plain=1\#L1-L4}{challenge32\_reason.adoc}) & 1\\ \cline{2-5}
& \newpolicy\label{p:11} & sitespeedio/\newline sitespeed.io & \enq{We \textbf{don't use ChatGPT to code} sitespeed.io but we \textbf{prompt it to write a blog post} about sitespeed.io as it was Steve Jobs writing it and it turned out quite good.} (\href{https://github.com/shred/acme4j/blob/ec726f6859b12ed59830e6e80a50daf5f034345c/CONTRIBUTING.md?plain=1\#L5-L15}{CONTRIBUTING.md}) & 4\\ \cline{2-5}
& \newpolicy\label{p:13} & theokanning/\newline openai-java &
  \enq{How to Contribute Add POJOs to API library I usually have \textbf{ChatGPT write them for me by copying and pasting} from the OpenAI API reference (example chat [link]), but double-check everything because \textbf{Chat always makes mistakes}, especially around adding `@JsonProperty` annotations.} (\href{https://github.com/TheoKanning/openai-java/blob/269096609cb81dad5e21c8d19e669a656bebacf4/CONTRIBUTING.md?plain=1\#L6}{CONTRIBUTING.md}) & 1\\
\midrule
\multirow{7}{*}[-2em]{\underline{\textbf{sup}}} 
& \newpolicy\label{p:8} & avaloniaui/\newline avalonia &
  \enq{Please provide a good description of the PR. Not doing so will delay review of the PR at a minimum, or may cause it to be closed. \textbf{If English isn't your first language, consider using ChatGPT or another tool to write the description.} If you're looking for a good example of a PR description see [PR link] for example.} (\href{https://github.com/AvaloniaUI/Avalonia/blob/2b3b1ef1e98a582785ad3dbe5810c466b0cfe472/CONTRIBUTING.md?plain=1\#L42}{CONTRIBUTING.md}) & 1\\ \cline{2-5}
& \newpolicy\label{p:9} & hardisgroupcom/\newline sfdx-hardis &
  \enq{Learn how to solve deployments errors that can happen during merge requests [...] SOS, I'm lost [...] - Call your release manager, he/she's here to help you! \textbf{Google / ChatGPT / Bard} the issue} (\href{https://github.com/hardisgroupcom/sfdx-hardis/blob/f7089a6bfbf4dbc1cfaebb3562d841d2fa892833/docs/salesforce-ci-cd-solve-deployment-errors.md?plain=1\#L50-L51}{salesforce-ci-cd-solve-deployment-errors.md}) & 1\\ \cline{2-5}

& \newpolicy\label{p:12} & spring-projects/\newline spring-cli &
  \enq{Large Language Models such at OpenAI's ChatGPT offer a powerful solution for generating code using AI. ChatGPT is trained not only on Java code but also on various projects within the Spring open-source ecosystem. Using a simple command, you can describe the desired functionality, and ChatGPT \textbf{generates a comprehensive `README.md` file} that provides step-by-step instructions to achieve your goal ... For further improvements and accuracy, you can get \textbf{ChatGPT to rewrite the description by using the --rewrite option}: The `ai add` command lets you add code to your project generated by using OpenAI's ChatGPT.} (\href{https://github.com/spring-attic/spring-cli/blob/7ea94f0dd246b829c96f1b11bec640c94b1760d2/docs/modules/ROOT/pages/ai-guide.adoc?plain=1\#L5-L9}{ai-guide.adoc}) & 5\\ 
\bottomrule
\end{tabularx}
\end{table*}

In the following, we present the results of our analysis of policies and usage guidelines and our developer survey.

\subsubsection{Policies and Usage Guidelines of GenAI Tools}

As mentioned above, Table~\ref{tab:reg} lists 13 software artifacts from 12 GitHub repositories that presented policies or usage guidelines for GenAI usage.
We classified them into prohibitive, restrictive, and supportive policies.

\textbf{Prohibitive:} Policies P\ref{p:1}--P\ref{p:6} illustrate community decisions that exclude GenAI usage in the projects.
Maintainers of \texttt{jqwik-team/jqwik} raised concerns related to the copyright situation around GenAI-generated content (P\ref{p:1} and P\ref{p:2}).
Similarly, maintainers of \texttt{shoelace-style/shoelace} addressed ethical and licensing issues arising from the inclusion of GenAI-generated code (P\ref{p:3}).
Regarding code reviews, maintainers of \texttt{katsutedev/mal4j} (P\ref{p:5}) and \texttt{shred/acme4j} (P\ref{p:6}) explicitly stated that contributions generated by GenAI are not acceptable.
The project \texttt{turms-im/turms} (P\ref{p:4}) discouraged the use of GenAI-generated responses in discussions, citing concerns over the lack of critical thinking and responsibility.
In addition, the maintainers proposed to incorporate indicators for identifying possible GenAI usage and suggested tool support, for example, a ChatGPT detector published on HuggingFace~\cite{DBLP:journals/corr/abs-2301-07597}.
These regulations demonstrate how open-source communities are beginning to establish boundaries and safeguards regarding GenAI tools in collaborative open-source software development.

\textbf{Restrictive:} Policies P\ref{p:7}--P\ref{p:13} restrict the use of GenAI tools in software development workflows without completely banning it.
For example, the maintainers of \texttt{graycoreio/daffodil} (P\ref{p:7}) require developers to disclose any use of GenAI as a prerequisite for submitting a pull request. Meanwhile, maintainers of \texttt{owasp/wrong-secrets}, \texttt{sitespeedio/sitespeed.io}, and \texttt{theokanning/openai-java} (P\ref{p:10}, P\ref{p:11}, and P\ref{p:13}) advised caution when using GenAI, warning of potential inaccuracies and security risks, such as inadvertent secret leakage due to the fact that tool vendors use prompts for reinforcement learning.

\textbf{Supportive:} Policies P\ref{p:8}--P\ref{p:12} outline supportive guidelines for the use of GenAI tools.
The maintainers of \texttt{avaloniaui/avalonia} (P\ref{p:8}) encouraged the use of GenAI to help draft pull request descriptions to support the code review process.
In \texttt{hardisgroupcom/sfdx-hardis} (P\ref{p:9}), maintainers recommended using GenAI for Q\&A support, particularly for troubleshooting deployment issues.
The project \texttt{spring-projects/spring-cli} (P\ref{p:12}) promoted the use of GenAI to generate \texttt{README.md} files and has even developed GenAI tooling to support automated documentation rewriting. 

Overall, these policies and usage guidelines reflect a growing awareness of both the opportunities and risks of GenAI tools in open-source software projects and the willingness of the maintainers to guide their usage. An interesting direction for future work is extending the analysis to cover more policies and guidelines and correlating them with other project characteristics. For example, one could hypothesize that GPL-licensed projects are more likely to have restrictive regulations.

\subsubsection{Developer Survey on GenAI Governance}

Based on the analysis of the before-mentioned policies and usage guidelines, we designed ten questions regarding (i) the necessity of GenAI tool guidance; (ii) the necessity of documenting prompts and their generated contents; (iii) actions on generated content before integrating; and (iv) the rationale behind policies and usage guidelines of real-world GenAI tools. In the following, we will use $D$ to refer to individual developers who participated in our survey.

\textbf{General GenAI Usage Guidance:}
Five developers highlighted the necessity of regulating the usage of GenAI tools in software projects. They cited concerns such as copyright issues, license violations, and ethical considerations as key reasons for establishing guidelines. For instance, respondent $D_3$ remarked that \enq{using GenAI is a highly ethical question. With a regulation, one can take a stance.}
The motivation for guidelines and regulations varied, with $D_6$ stating that \enq{it [GenAI tool usage] is convenient, but can be detrimental to the codebase if used fully unregulated,} while $D_2$ noted that \enq{it largely depends on the risk appetite and sensitivity of the project/organization.}
Interestingly, $D_5$ expressed a negative view of regulating GenAI tool usage, arguing that it could hinder productivity. They stated:
\enq{No, instead, humanity must fully harness the potential of AI to unleash productivity. Regulating its usage too tightly would hinder innovation and slow down progress. Instead of imposing external regulations on AI usage, human society should develop autogenous forms of regulation, driven by shared values, ethical guidelines, and adaptive practices.}
When asked about specific aspects of software projects that should be regulated, developers expressed concerns primarily about unlicensed training datasets and potential licensing conflicts associated with AI-generated code. For example, $D_1$ observed \enq{It's unclear what the license of AI-generated code is. AIs have been trained on all kinds of licenses, so what license is the generated code?}

\textbf{Documenting Prompts and Generated Content:}
$D_{5,6,8}$ emphasize the importance of documenting prompts and generated content to ensure accountability in software projects. They suggest two methods to achieve this: (1) associating prompts with their functionality and sharing them under a \emph{CC BY 4.0} license, and (2) embedding prompts as code comments or in project documentation, supplemented by shared conversations, e.g., via ChatGPT links.
Despite some developers considering prompt documentation unnecessary, the majority agreed that it is valuable to understand the extent of GenAI's contributions to a project.
This documentation is essential for assessing which code is potentially affected by copyright and licensing issues; it might also prove useful for later maintenance activities.

\textbf{Actions on Generated Content Before Integration:}
Developers are, compared to manually written code, more likely to perform code reviews and license compliance checks on AI-generated content before integrating it into their projects.
Three developers highlighted these practices as crucial steps to ensure the quality and compliance of GenAI-based contributions.
Additionally, some developers indicated that they rely on automated tools, e.g., code quality checks or automated testing, to evaluate generated content.
One developer noted the importance of adding comments to the document generation context.
Interestingly, $D_2$ explicitly stated that no additional actions are necessary, explaining that \enq{all content in the PR will be subjected to rigorous review and testing regardless.}
This response reflects the perspective that standard testing and code review are sufficient to ensure the quality of both AI-generated and manually created content.

\textbf{Project-Specific GenAI Usage Guidance:}
The feedback we received from open-source developers regarding GenAI tool guidance reflects a combination of ethical, legal, and practical considerations.
For example, the project owner of \texttt{jqwik-team/jqwik} ($D_3$) described their decision to disallow the use of GenAI tools as an \enq{ethical decision due to all its collateral damages.}
This statement suggests a strong position against the potential implications of accepting AI-generated contributions, with a particular focus on copyright and ethics.
The regulation in the accompanying contributor agreement (P\ref{p:1}) explicitly prohibits contributions created using GenAI tools, citing the unresolved legal implications of training AI models on public code repositories.
Similarly, the project owner of \texttt{owasp/wrongsecrets} ($D_2$) focuses on the ethical risks of using GenAI when describing the rationale behind guideline P\ref{p:10}.
They highlight the importance of vigilance when handling sensitive data, particularly in the \texttt{wrongsecrets} project.
They reported: \enq{This is a recommendation meant for people using WrongSecrets, and it applies more broadly than WrongSecrets or even OWASP itself. You should be conscious about what data you share, and be vigilant that you don't input sensitive data, since tenant boundaries are murky at best.}
This raises a broader concern about how user input may be stored or reused by GenAI systems. The associated recommendation emphasizes that GenAI tools often rely on reinforcement learning, which could expose sensitive data to unintended parties.

Guideline P\ref{p:7}, which requires the disclosure of AI usage in pull requests, received support from three developers ($D_4$, $D_5$, $D_6$).
$D_4$ emphasized that disclosure depends on whether a \enq{key idea} was generated by AI, while $D_5$ highlighted the importance of transparency for license compliance.
$D_6$ added that disclosing the percentage of GenAI involvement in contributions could reduce the likelihood of generated \enq{noise PRs} and improve code review efficiency.
This reflects a growing recognition of the need for transparency in collaborative software development, where understanding the role of AI in contributions can improve accountability and ensure compliance.

Opinions diverge considerably for P\ref{p:3}, which prohibits AI-generated code.
$D_4$ opposed such restrictions, viewing them as unnecessary limitations that could stifle productivity and innovation.
$D_6$ criticized the policy as being overly cautious, suggesting that asking contributors to \enq{disclose percentage} of generated content is sufficient.
$D_7$ supported the regulation, noting its alignment with their own concerns about the ethical and legal implications of using GenAI tools.
$D_5$, pointed to \enq{ethical and legal ambiguities related to AI-generated code}, describing them as \enq{maintainer's main concerns.}
They specifically highlighted that \enq{AI models are likely trained on large datasets that include open-source codebases with various licensing terms.}
$D_4$ and $D_6$'s feedback on P\ref{p:4}, which regulates the use of AI in community discussions, emphasizes concerns about the high false positive rates of AI identification tools and warns against deferring critical decisions to automation.
$D_5$ argued that while LLMs are suitable for repetitive tasks and generic translations, they lack the creativity needed for meaningful contributions.
This aligns with the cautionary tone of the regulation, which warns about overreliance on AI-generated content.

\begin{summarybox}{RQ2}
We found 13 policies and guidelines on GenAI usage in open-source software projects, including strict policies prohibiting GenAI usage, policies requiring attribution, but also guidelines encouraging contributors to use GenAI, for example, for translating natural language text.
The results of our developer survey reflect the tension between anticipated productivity gains of GenAI tools and legal and ethical implications of their usage.
\end{summarybox}

\section{Impact of GenAI Usage on Code Churn}

The goal of \textbf{RQ3} was to examine the impact of GenAI usage on open-source software projects.

\subsection{Method}

The GenAI mentions we identified as part of \textbf{RQ1} allow us to approximate the point in time when the 156 open-source projects in our sample started using the GenAI tools.
We included 151 repositories with true positive GenAI mentions that did not prohibit the use of GenAI tools. That is, we excluded five repositories prohibiting GenAI usage according to our \textbf{RQ2} analysis.
Hence, we use self-admitted GenAI usage as a proxy for GenAI tool adoption.

To assess the impact of GenAI tool usage, we calculated the code churn, as defined in the GitClear report (see Section~\ref{sec:introduction}), before and after the first self-admitted GenAI usage.
Code churn is a widely recognized indicator of software maintainability~\cite{munson1998code}.
Churn rates can signal challenges such as higher defect density~\cite{nagappan2005use}; files affected by technical debt tend to be more defect-prone~\cite{wehaibi2016examining}.
Code churn is particularly relevant for understanding the maintainability of LLM-generated source code, which might introduce redundancies or bugs that result in changes soon after adding generated code.

The specific notion of code churn introduced by GitClear measures whether added or modified code is updated again within 14 days of the initial commit.
Therefore, it serves as an indicator of the maturity of the code that developers add or modify.
The 2024 GitClear report~\cite{gitclear2024} suggested that code churn has been continuously increasing since the adoption of GenAI tools in software projects.

To answer \textbf{RQ3}, we selected repositories with at least one self-admitted GenAI usage.
The first recorded GenAI mentions in the commit history served as the adoption point ($t_\text{mention}$).
Code churn was analyzed across two timeframes:

{\small
\begin{itemize}
    \item pre-GenAI adoption: The 360 days preceding $t_\text{mention}$
    \item post-GenAI adoption: The 360 days following $t_\text{mention}$
\end{itemize}}

In the following, the term \textit{churned lines} refers to the number of lines that were added or modified within the defined timeframes (pre-GenAI adoption or post-GenAI adoption).
For each commit, we track the changes introduced with the commit and whether those changes were modified again within a 14-day~window.

Specifically, we defined code churn as the percentage of lines that are reverted or updated within 14 days after they were initially added or modified.
We added a second definition that focuses on churned files instead of lines to gain a more comprehensive understanding of the impact of GenAI adoption on the selected repositories.

\textbf{Line-based churn} measures the percentage of lines (1) that the commit added or modified and (2) that were changed again within 14 days after the commit. This metric captures the frequency with which individual lines are churned, indicating potential code maintainability challenges. Line-based churn $ch_L$ for a commit $c$ is defined as:

{\small
\begin{equation*}
ch_L(c) =  
\frac{\text{\#lines changed again within 14 days}}{\text{total \#lines changed by } c}
\end{equation*}}

\textbf{File-based churn} measures the percentage of files (1) that the commit added or modified and (2) that were changed again within 14 days after the commit.
For this definition, we consider all changes to the files, regardless of the specific lines that were changed.
File-based churn $ch_F$ for a commit $c$ is defined as:

{\small
\begin{equation*}
ch_F(c) =  
\frac{\text{\#files changed again within 14 days}}{\text{total \#files changed by } c}
\end{equation*}}

For each granularity level ($ch_L$, $ch_F$), to understand trends, we report changes in the average code churn over multiple commits. We calculated:
{\small
\begin{enumerate}
\item The average churn per repository, comparing pre- and post-GenAI adoption using \emph{Wilcoxon signed-rank test}~\cite{wilcoxon1945individual}. We applied the \emph{Wilcoxon Z statistic} $r$ to measure the paired effect and interpreted the effect size as follows~\cite{cohen2013statistical}:  $|r| < 0.1$ as \textit{negligible}, $0.1 \leq |r| <0.3$ as \textit{small}, $0.3 \leq |r| < 0.5$ as \textit{medium}, and  $0.5 \le |r|$ as \textit{large};
\item The average churn over all commits in all repositories, comparing pre- and post-GenAI adoption using \emph{Mann-Whitney test}~\cite{mann1947test}. We applied the Cliff $\delta$~\cite{cliff1993dominance} to measure the independent effect and interpreted the effect size as follows~\cite{romano2006exploring}:  $|\delta| < 0.147$ as \textit{negligible}, $0.147 \leq |\delta| <0.33$ as \textit{small}, $0.33 \leq |\delta| < 0.474$ as \textit{medium}, and  $0.474 \le |\delta|$ as \textit{large}.
\end{enumerate}}


\begin{table}[tb]
    \centering
    \caption{Effect size of significant code churn differences pre- vs. post-GenAI adoption, measured using Wilcoxon signed-rank test ($\alpha = 0.05$) and Wilcoxon Z statistic $r$ ($n=151$).}
    \label{tab:wilcoxon}
    \begin{threeparttable}
    \begin{tabular}{llrrr}
    \toprule
Churn Type &  Effect size & \#Significant & Sum sig. & Not sig. \\
\midrule
\multirow{4}{*}{File-based} & negligible& \textcolor{orange}{5} | \textcolor{cyan}{10} & 15 & 19 \\
& small & \textcolor{orange}{14} | \textcolor{cyan}{30} & 44 & 8 \\
& medium & \textcolor{orange}{9} | \textcolor{cyan}{23} & 32 & 1 \\
& large& \textcolor{orange}{2} | \textcolor{cyan}{26} & 28 & 4 \\
\cline{2-5}
& sum & \textcolor{orange}{30} | \textcolor{cyan}{89} & 119 & 32\\
\midrule
\multirow{4}{*}{Line-based} & negligible & \textcolor{orange}{5} | \textcolor{cyan}{5} & 10 & 22 \\
& small & \textcolor{orange}{13} | \textcolor{cyan}{33} & 46 & 7 \\
& medium & \textcolor{orange}{7} | \textcolor{cyan}{24} & 31 & 0 \\
& large & \textcolor{orange}{8} | \textcolor{cyan}{25} & 33 & 2 \\
\cline{2-5}
& sum & \textcolor{orange}{33} | \textcolor{cyan}{87} & 120 & 31 \\
    \bottomrule
\end{tabular}
\begin{tablenotes}
      \footnotesize
      \item Each value corresponds to the number of repositories exhibiting an \textcolor{orange}{increasing trend} or a \textcolor{cyan}{decreasing trend}, respectively.
\end{tablenotes}
\end{threeparttable}
\end{table}

\begin{table}[tb]
\caption{Distribution of code churn patterns based on RDD ($\alpha=0.05$, $n=149$).}
\label{tab:rdd}
\resizebox{\columnwidth}{!}{
\begin{tabular}{llrrrr}
\toprule
\multirow{2}{*}{Churn Type} & \multirow{2}{*}{Trend} & \multicolumn{2}{c}{Slope} & \multirow{2}{*}{Sum sig.} & \multirow{2}{*}{Not sig.} \\ 
&& \#Positive & \#Negative & & \\
 \midrule
 \multirow{2}{*}{File-based} & Upward & 3 (11.5\%) &   12 (46.2\%) & \multirow{2}{*}{26} & \multirow{2}{*}{123} \\
& Downward  & 4 (15.4\%) & 7 (26.9\%) & & \\
\midrule
\multirow{2}{*}{Line-based} & Upward  & 5 (16.7\%) & 10 (33.3\%) & \multirow{2}{*}{30} & \multirow{2}{*}{119} \\
& Downward & 3 (10.0\%) & 12 (40.0\%) &  & \\
\bottomrule
\end{tabular}
}
\end{table}

We further used a \emph{Regression Discontinuity Design} (RDD)~\cite{thistlethwaite1960regression, imbens2008regression} to study the impact of GenAI adoption on code churn.
RDD is a quasi-experimental method evaluating the impact of an intervention by comparing outcome data points before and after a cutoff point (in our case, the first GenAI mention in a repository). 
This method has been applied in software engineering before, for example, to assess the impact of introducing code review bots and GitHub Actions to software repositories~\cite{wessel2020effects, wessel2023github}. 

We categorized the patterns that emerged from the RDD analysis based on two characteristics: (1) \emph{trend} and (2) \emph{slope}. The \emph{trend} characteristic captures whether code churn increases or decreases from the pre- to the post-GenAI adoption period. The \emph{slope} characteristic captures whether and how the rate of change in the trend line shifts at the point of GenAI adoption.
The Ordinary Least Squares (OLS) model used in the RDD analysis requires a minimum number of observations to estimate its four parameters (intercept, time trend, treatment effect, and interaction) while maintaining positive degrees of freedom~\cite{imbens2008regression}. To meet this requirement, we aggregated commit-level churn values into weekly data points and fit the RDD model to the resulting weekly time series. For each repository, we required a minimum time span of five weeks, each containing at least one commit, as well as at least one such week on either side of the GenAI adoption point, ensuring that both pre- and post-adoption trends could be estimated. After applying these thresholds, we excluded two repositories with insufficient data.

\begin{figure*}[tb]
    \centering
\includegraphics[width=0.9\textwidth]{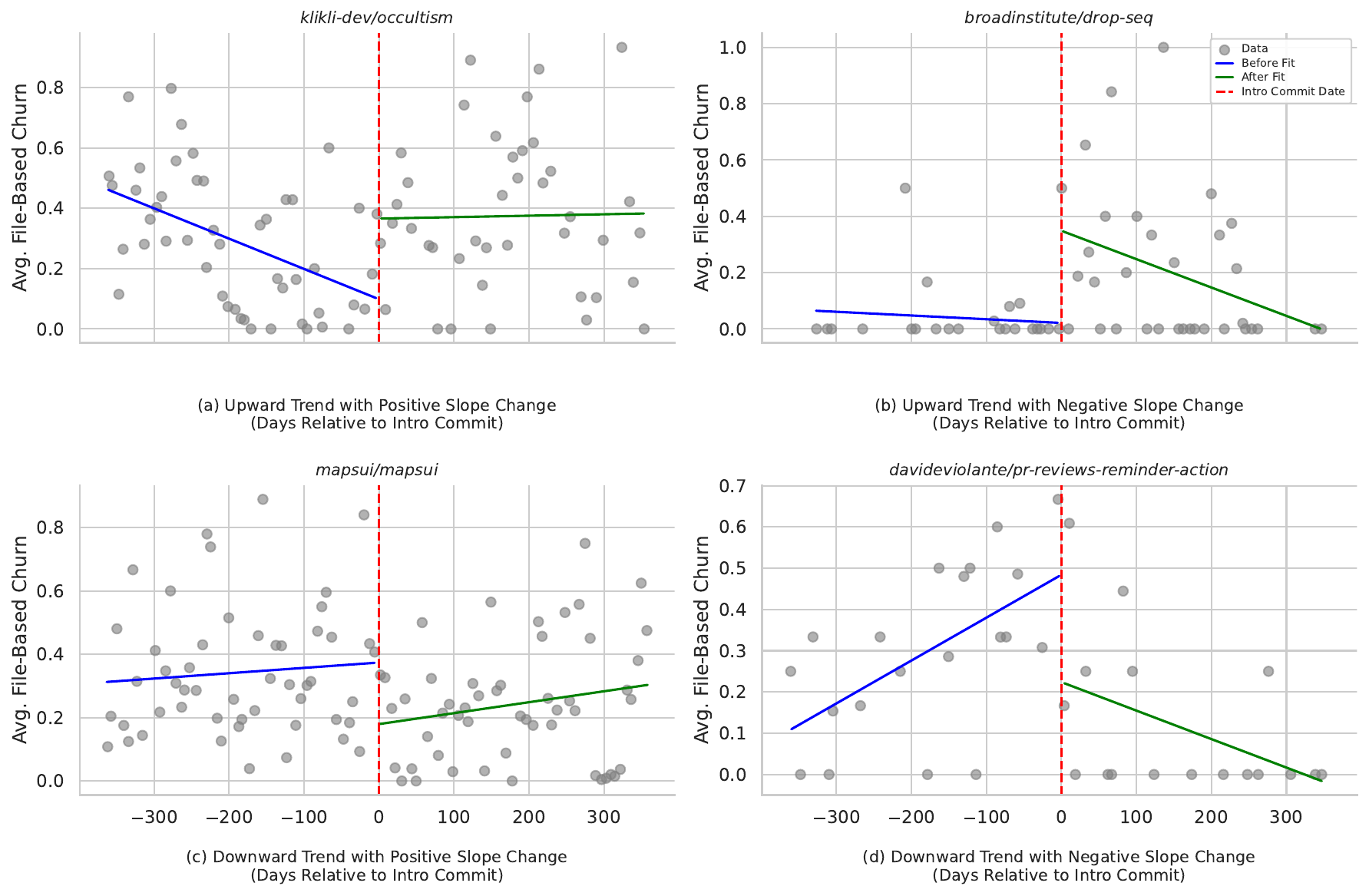}
    \caption{One representative repository for each of the four RDD patterns of file-based churn, showing average file-based churn per week (gray), pre-adoption fit (blue), post-adoption fit (green), and the first GenAI mention (red dashed line).}
    \label{fig:example}
\end{figure*}

For the 149 repositories included, we identified four patterns, that we describe in the following: 
{\small
\begin{itemize}
\item[\textbf{a.}] \textbf{Upward trend with positive slope change:} This pattern shows code churn increasing after GenAI adoption with an increasing rate of change, which means that the churn grows progressively faster.
\item[\textbf{b.}] \textbf{Upward trend with negative slope change:} Here, the code churn increases after GenAI adoption, but the rate of increase decelerates over time, suggesting that the initial churn increase gradually stabilizes over time.
\item[\textbf{c.}] \textbf{Downward trend with positive slope change:} In this pattern, churn decreases after GenAI adoption, but the change rate slows down over time.
\item[\textbf{d.}] \textbf{Downward trend with negative slope change:} This pattern exhibits decreasing churn after GenAI adoption with an accelerating rate of decline, which means that the churn reduction progressively increases.
\end{itemize}}
These patterns (with examples illustrated in Figure~\ref{fig:example}) provide a useful framework for analyzing how code churn metrics change after GenAI adoption in different project contexts.
To assess the robustness and causal interpretability of the RDD-based patterns, we conducted robustness checks. First, in addition to each repository’s project-specific adoption point (i.e., the first commit explicitly reporting GenAI assistance), we estimated parallel RDD models using two global cutoff dates corresponding to major GenAI releases: GitHub Copilot (June 2021) and ChatGPT (November 2022). 
Second, we assessed robustness to bandwidth selection by estimating the RDD under multiple symmetric temporal windows around the cutoff (±90, ±180, and ±360 days), while keeping all other model specifications constant. Third, we conducted placebo tests by shifting each repository’s cutoff date forward and backward in time and re-estimating the same models. Across all robustness checks, the results remained stable, indicating that our main RDD findings do not depend on arbitrary cutoff choices or temporal window specifications.

Finally, to examine heterogeneity across different forms of GenAI usage, we further disaggregated the RDD analysis by the GenAI-assisted tasks identified in \textbf{RQ1}  (e.g., generation, optimization, and maintenance). We additionally performed a manual verification of every subcode assigned to each task in Table~\ref{tab:task}, including those subcodes likely to influence code churn. The full set of GenAI generation task categories includes code, test data, comment, test file, regex, test method, Zod schema, test class, and configuration. GenAI optimization task categories encompass all subcodes of code refactoring and code improvement. GenAI maintenance task categories include comment revision, color suggestions, dependency upgrades, and version updates. 

   \begin{table}[tb]
    \centering
    \caption{Significant RDD churn discontinuities by task category (Generation/Optimization/Maintenance).}
    \label{tab:task_churn_summary}
    \begin{threeparttable}
    \begin{tabular}{llrrr}
    \toprule
Churn Type & Task & \#Significant & Sum sig. & Not sig. \\
\midrule
\multirow{3}{*}{File-based}
& Generation   & \textcolor{orange}{10} $|$ \textcolor{cyan}{6} & 16 & 68 \\
& Optimization & \textcolor{orange}{1}  $|$ \textcolor{cyan}{1} & 2  & 12 \\
& Maintenance  & \textcolor{orange}{0}  $|$ \textcolor{cyan}{1} & 1  & 2  \\
\midrule
\multirow{3}{*}{Line-based}
& Generation   & \textcolor{orange}{9}  $|$ \textcolor{cyan}{6} & 15 & 69 \\
& Optimization & \textcolor{orange}{2}  $|$ \textcolor{cyan}{2} & 4  & 10 \\
& Maintenance  & \textcolor{orange}{0}  $|$ \textcolor{cyan}{2} & 2  & 1  \\
    \bottomrule
\end{tabular}
\begin{tablenotes}
      \footnotesize
      \item Each value corresponds to the number of repositories exhibiting an \textcolor{orange}{increasing trend} or a \textcolor{cyan}{decreasing trend}, respectively.
\end{tablenotes}
\end{threeparttable}
\end{table}

\subsection{Results}

Table~\ref{tab:wilcoxon} illustrates the variations in code churn of the studied repositories.
Of the 151 repositories with self-admitted GenAI usage, 119 had a significant difference in file-based churn, and 120 had a significant difference in line-based churn ($p < 0.05$).
Eleven repositories had an increasing file-based churn with a medium-to-large effect size, and 15 had an increasing line-based churn with a medium-to-large effect size.
A decreasing churn was more common: 49 repositories had a decreasing file-based churn with a medium-to-large effect size, and 49 repositories had a decreasing line-based churn with a medium-to-large effect size.

Besides the average code churn per repository pre- and post-GenAI adoption, we also compared the average code churn over all commits in our dataset pre- and post-GenAI adoption.
The average file-based code churn decreased from 0.17 to 0.06 with a significant difference ($p < 0.05$) and a medium effect ($|\delta| = 0.42$), the average line-based churn decreased from 0.68 to 0.50 with a significant difference ($p < 0.05$) and a negligible effect ($|\delta| = 0.09$).
These results are contrary to our expectations because the GitClear report was very bold in claiming that code churn increased for the projects they studied, suggesting a \enq{downward pressure on code quality}~\cite{gitclear2024}.
While we observed that some repositories have an increasing trend in code churn, both the overall trend and the trend in many individual repositories point to a decreasing code churn over time.
Therefore, with our data and methodology, we cannot confirm this claim.

Table~\ref{tab:rdd} summarizes the results of our RDD analysis.
We observed that only 26 (file-based) and 30 (line-based) repositories showed significant code churn trends ($p < 0.05$).
For file-based churn, an overall upward trend with a negative slope after the cutoff date was most common (12 repositories).
For line-based churn, an overall downward trend with a negative slope was most common (12 repositories).
However, there were almost as many repositories (10) with an overall downward trend, but a positive slope after the cutoff date.
We found 15 repositories with a significant upward trend in file-based churn and 15 with a significant upward trend in line-based churn.
However, most of them had a negative slope.
In addition, 11 projects had a significant downward trend in file-based churn and 15 had a significant downward trend in line-based churn.
We cannot conclude that there is a general trend toward increasing code~churn.

To deepen our understanding of which GenAI-assisted tasks (\textbf{RQ1}) are most strongly associated with churn changes, we have further disaggregated the RDD results by task (see Table~\ref{tab:task_churn_summary}). Across all models, significant discontinuities are concentrated in the generation category, including the generation of code, test data, test methods, regular expressions, and comments. In contrast, optimization and maintenance tasks do not show consistent patterns. In summary, this task-level analysis reveals that GenAI-assisted generation tasks appear more rework-prone than other forms of assistance.

\begin{summarybox}{RQ3}
Our results revealed that for most of the repositories analyzed, there was no significant change in code churn after GenAI adoption.
We did find 15 repositories with an overall upward trend in line-based code churn after the first GenAI mention.
However, for most of them, the slope was negative.
In addition, we also found 15 repositories with a downward trend.
Among the GenAI tasks identified in \textbf{RQ1}, generation tasks show a stronger impact on code churn than other tasks.
These results indicate that more research is required to understand why certain projects or GenAI tasks are affected and others are not, and how higher (or lower) code churn relates to the long-term maintainability of software projects.
\end{summarybox}

\section{Discussion}

In this section, we discuss and contextualize the results of our three research questions and summarize the implications for software developers and researchers.

\subsection{RQ1: Reasons for Mentioning GenAI Tools}

By focusing on self-admitted GenAI usage, that is, explicit mentions of GenAI tools in source code comments, commit messages, and documentation files, we gained a thorough understanding of how and why developers acknowledge GenAI tools in open-source projects.
One central contribution of this paper is our taxonomy of assisted tasks, content types, and usage purposes (see Tables~\ref{tab:task}, \ref{tab:content}, and \ref{tab:purpose}).

\textbf{Our taxonomy provides a multi-dimensional characterization (task, content type, purpose), where prior studies only categorize tasks assigned to GenAI tools.}
Our analysis revealed that developers primarily use GenAI tools for code generation, natural language translation, and code refactoring. Tufano et al.~\cite{Tufano2024} explored mentions of ChatGPT in commits, PRs, and issues. They identified that the three most common task categories were feature implementation and enhancement, software quality, and documentation.
In our study, we present a more fine-grained and comprehensive categorization of tasks automated by both ChatGPT and GitHub Copilot.
For studies targeting GitHub repositories, it is crucial to consider GitHub Copilot as well, because (1) unlike ChatGPT, it is a tool tailored to software development, and (2) it is more deeply embedded in developers' workflows (their local editors, but also into the GitHub platform as a whole).
Moreover, we complement the task categories by specifically discussing content and usage purposes.
In addition, we identified patterns of human intervention.
Hou et al.~\cite{hou2023large} reviewed literature on LLMs for software engineering.
They found that software engineering research has a strong focus on code generation and program repair.
We complement this observation with a detailed taxonomy of how open-source developers use LLM-based tools in their projects.
In addition to code generation, we found that internationalization and natural language translation are common use cases for LLMs in open-source software projects (49 instances, see Table~\ref{tab:task}). This finding is aligned with Tufano et al.~\cite{Tufano2024}, who identified 12 instances of ChatGPT usage for internationalization. Such use cases highlight the need to support not only programming tasks but also broader software development activities.
We further found instances of projects regulating the usage of GenAI tools, which we analyzed in more detail as part of \textbf{RQ2}.
While our study partially confirms previous studies on software development tasks being automated using GenAI tools, we contribute three novel perspectives: (1) some developers deeply care about acknowledging GenAI usage in open-source software, (2) open-source maintainers try to actively guide and regulate GenAI usage, and (3) issues with generated code can trigger human interventions in open-source software projects.

Based on our findings, we recommend that \textbf{researchers} investigate developers' rationale behind self-admitted (vs. hidden) GenAI usage. Our notion of self-admitted GenAI usage, inspired by self-admitted technical debt~\cite{potdar2014exploratory}, can be a valuable lens for studying GenAI usage in practice.
Of course, only a fraction of the generated software artifacts contain GenAI mentions, and the artifacts that are documented might not be representative of the overall GenAI usage. Better understanding when and why developers decide to self-admit GenAI usage is one potential direction for future work.
Moreover, our annotated dataset of self-admitted GenAI usage can serve as a starting point to build a tool to automatically identify true positive GenAI mentions according to the definition presented in Section~\ref{sec:identifying-mentions}. An improved and scaled detection of self-admitted GenAI usage would allow researchers to build larger datasets that could then enable more comprehensive studies on code quality and maintainability of generated code.

\textbf{Software developers} can browse our taxonomy of tasks, content types, and purposes to identify potential applications of GenAI tools in their projects. One central aspect is whether to establish guidelines clarifying in which cases project maintainers require contributors to disclose and acknowledge GenAI usage (see also Section~\ref{sec:discussion-rq2}).

We found that some acknowledgments were combined with warnings about the potential negative implications of GenAI tool usage.
Sometimes, GenAI tools were also blamed for issues.
However, acknowledgment can also serve a positive purpose, e.g., documenting prompts.
The question arises not only when to acknowledge GenAI usage, but also which context to document beyond the tool name (which we focused on).
In which cases does it make sense to document complete prompts and where and how should one document the generation context? Such questions can be addressed both from a \textbf{scientific} and from a \textbf{practical} perspective.
A more standardized approach for documenting GenAI contributions is required, since most self-admitted GenAI usages did not document the generation context beyond brief summaries.
As discussed above, we found cases documenting prompts behind GenAI usage. \textbf{GenAI tool builders} (e.g., of IDE plugins) could offer an option to record these contexts, enabling transparent acknowledgment of GenAI usage and prompt reuse.

\subsection{RQ2: Existing Guidelines for GenAI Usage}
\label{sec:discussion-rq2}
Motivated by the purpose categories \emph{Guidance and Best Practices} that we identified while answering \textbf{RQ1} (see Table~\ref{tab:purpose}), we further explored the policies and usage guidelines for GenAI tools that we found (see Table~\ref{tab:reg}).
Their content ranged from encouraging developers to use GenAI tools to prohibiting their usage entirely. 

\textbf{Our developer survey suggests a broad spectrum of positions that cover ethical, legal, and practical considerations.}
Mentioned aspects include the unclear copyright situation of the training data, the unclear implications for generated content, data privacy risks when sharing inputs with GenAI systems, and concerns regarding code quality and maintainability.
Moreover, a majority of our survey participants agreed that the regulation of GenAI usage is necessary in open-source projects.
In relation to that, participants argued for transparent disclosure of GenAI usage and also for documenting the generation context. It is unclear how much transparency is required and what purposes it can serve:
Is a binary flag sufficient? Or is it better to document the percentage of generated content, as suggested by a participant? Or the whole prompt? Do only manually written prompts need to be disclosed, or also system prompts?
This aspect is aligned with the questions raised in the discussion for \textbf{RQ1} about prompt context.

Our results suggest that \textbf{software developers}, especially those maintaining open-source software projects, should \textbf{articulate a clear position} regarding GenAI usage in their projects.
The spectrum of possible positions ranges from a general recommendation to use GenAI tools, over recommendations for specific tools and use cases, to more restrictive policies requiring an extensive peer review of generated content, or policies prohibiting GenAI usage completely.
Open-source projects should \textbf{clearly communicate expectations} regarding GenAI usage to their contributors.
For downstream consumers of open-source dependencies, explicit GenAI policies serve as a signal of due diligence that may influence their dependency selection.

Our analysis of policies, guidelines, and developers' positions regarding GenAI regulation provides a solid foundation for \textbf{researchers} to design and conduct further \textbf{studies on how software projects regulate} GenAI usage and how such regulations impact development activity.
An idea worth exploring is whether existing GenAI tools could be augmented to \textbf{capture provenance information during generation} that could be automatically added to source code comments, commit messages, or other artifacts such as Software Bills of Materials (SBOMs)~\cite{DBLP:journals/computer/Riehle25} or Software Bill of Materials for AI (SBOM for AI)~\cite{xia2023aibom}.
There are already open-source projects that extensively document prompts in commit messages.\footnote{\href{https://github.com/cloudflare/workers-oauth-provider/commit/adcbb5de9c24f5b6a7dbea2e0a313a87c304d9bb}{github.com/cloudflare/workers-oauth-provider/commit/adcb...}} 
This provenance information is essential to study the \textbf{long-term impact of code generation on maintainability}, but also to support software supply chain transparency and vulnerability management.
Researchers can contribute to the \textbf{development of standardized metadata formats} to capture provenance and traceability information of code and other software artifacts.

\subsection{RQ3: Impact of GenAI Usage on Code Churn}

\textbf{Our results for \textbf{RQ3} challenge popular narratives about the impact of GenAI on software development.}
Contrary to claims in the GitClear report, which was extensively discussed in the software development community~\cite{gitclear-hackernews, gitclear-reddit}, we did not find an increasing code churn after GenAI adoption.
The overall trend we observed pointed in the opposite direction, i.e., we noticed a decreasing average code churn.
This is in line with a study by Grewal et al.~\cite{10.1145/3643991.3645072} which examined how ChatGPT-generated code is integrated into
GitHub projects.
They found that approximately 54\% of the generated code lines were integrated and only 2.5\% of the generated snippets were later modified.
However, our RDD analysis revealed that three repositories (3/26, 11.5\%) had a significant upward file-based code churn trend with a positive slope ($p < 0.05$). This indicates that a subset of repositories exhibited a progressively faster increase in code churn after GenAI adoption. To contextualize this finding, we note that a larger proportion of repositories (12/26, 46.2\%) showed an upward trend with a negative slope, i.e., churn increased, but the increase decelerated over time. In addition, 26.9\% (7/26) showed downward trends. This heterogeneity suggests that GenAI's impact on code churn is context-dependent rather than uniformly negative, as suggested by the GitClear report.

Our \textbf{RQ3} results motivate us to recommend that \textbf{researchers} explore the factors that contribute to increased code churn. The patterns we identified using our RDD analysis are a \textbf{valuable lens for clustering projects}.
This clustering can inform a detailed qualitative study of projects that exhibit similar patterns.
The difference between our results and the GitClear report can be partially attributed to the methodological differences between the studies.
While GitClear used a global cutoff date, we used the first GenAI mention in a repository as a proxy for GenAI adoption, thus following a more fine-grained approach.
Moreover, we introduced \textbf{file-level and line-level code churn} and calculated churn at the project-level and globally. Our definitions and the code in our supplementary material enable other researchers to consider code churn in their own studies.

For \textbf{software developers}, we further suggest \textbf{monitoring the impact of GenAI} on the software projects they maintain or contribute to. Our results suggest that the impact of GenAI adoption on the development activity in software projects might not be as clear as suggested by the GitClear report. Considering that we did notice an increasing code churn in several projects, it is nevertheless important for project maintainers to \textbf{monitor the development activity} and the quality of contributions. Going forward, we might develop our code churn implementation into a tool that project maintainers can easily integrate into their repositories.

\section{Related Work}

To situate our work, we organize related work into three themes that align with the dimensions explored in our study: 
(i) studies examining GenAI tasks and purposes, 
(ii) studies on risks and integration concerns around GenAI adoption, and 
(iii) studies on the impact of GenAI on software development processes.

\subsection{GenAI Tasks and Purposes}

Many researchers have focused on understanding how developers use GenAI tools across different software engineering activities and the types of content these tools generate. 

Besides our work and that of Tufano et al.~\cite{Tufano2024}, a few other studies have also established taxonomies of GenAI tasks in software development.
Sagdic et al.~\cite{10.1145/3643991.3645080} used semantic modeling and expert analysis to understand the topics developers discuss when interacting with ChatGPT, revealing 17 topics in seven categories, with over one-quarter of prompts focused on seeking programming guidance. 
Champa et al.~\cite{10.1145/3643991.3645077} defined 12 categories of software development tasks based on a literature review and applied these categories to analyze developers' interaction with ChatGPT. 
They found that code quality management and commit issue resolution represent the most frequent assistance requests.
These additional taxonomies provide further evidence of the breadth of software engineering activities in which developers rely on GenAI assistance. 

Research examining the purposes and contexts of GenAI usage has revealed several patterns in the ways developers integrate AI tools into their workflow.
Using the DevGPT dataset~\cite{10.1145/3643991.3648400}, Jin et al.~\cite{10.1145/3643991.3645074} found that LLM-generated code was rarely used as production-ready code, providing concrete evidence of the gap between GenAI capabilities demonstrated in research settings and their practical application in real-world development scenarios.
Their analysis revealed distinct purposes for AI-generated content: nearly one-third of the generated code was not integrated at all, whereas approximately one-quarter was incorporated into auxiliary files, such as README documentation files and test cases, rather than production codebases. 
This pattern suggests that developers may primarily use GenAI for explanatory and educational purposes rather than direct code production. 
Xiao et al.~\cite{xiao2024generative} studied GenAI-developer collaboration through the analysis of over 18K pull requests where descriptions were crafted by GitHub Copilot.
They found that developers complement AI-generated content with manual input, underlining the collaborative nature of human-AI interaction in producing development artifacts that require iterative refinement and enhancement.
Our analysis complements these studies by focusing on self-admitted GenAI usage, examining how and why developers explicitly acknowledge AI assistance in their development artifacts across different tasks and content types.

Despite the increasing amount of research studying GenAI assistance in software development, an important gap remains in our understanding of self-admitted GenAI usage patterns in the wild, particularly regarding how developers openly acknowledge and document GenAI assistance across different software engineering tasks and purposes.

\subsection{GenAI Risks and Integration Concerns}

The integration of GenAI tools into software development workflows has raised serious concerns regarding security risks and responsible adoption practices. 
Research in this area has focused on understanding the varied challenges developers face when incorporating these tools, ranging from immediate security and quality concerns to broader organizational and workflow integration issues.

Regarding security concerns, Sandoval et al.~\cite{sandoval2023lostcuserstudy} examined the security implications of using AI-written code assistants and found that LLMs may inadvertently introduce vulnerabilities into codebases, highlighting the need for careful screening when integrating AI-generated code. 
Asare et al.~\cite{asare2023ese} compared the performance of GitHub Copilot with human developers in secure coding tasks. They found that the GenAI tool exhibits patterns of security weaknesses similar to those of human programmers, raising questions about code review practices and security governance.

Code quality issues have emerged as another major risk factor closely related to security concerns.
Siddiq et al.~\cite{10.1145/3643991.3645071} used the DevGPT dataset to assess the quality of ChatGPT-generated code and found that such code suffers from issues including undefined variables, improper documentation, and security vulnerabilities related to resource management. 
These quality concerns extend across different programming contexts, as demonstrated by Moratis et al.~\cite{10.1145/3643991.3645070}, who analyzed 144 JavaScript code blocks generated by ChatGPT and found that approximately one-quarter of AI-written code blocks contained one or more violations. They observed that approximately 50\% of the violations related to best practices, 37\% related to code style issues, and 12\% were classified as error-prone violations.
Quality concerns increase when considering code modification versus creation. 
Rabbi et al.~\cite{10.1145/3643991.3645076} analyzed 1,756 AI-generated Python code snippets, systematically distinguishing between code created from scratch and modified code. They found that code modified using ChatGPT more frequently suffers from quality issues compared to ChatGPT-generated code. 
This pattern suggests that different types of AI assistance may require different governance approaches. 
Zhang et al.~\cite{10.1145/3643991.3645079} identified code smells in Kubernetes manifest files generated by AI tools, showing that quality concerns extend beyond traditional programming tasks to infrastructure-as-code artifacts.

The successful adoption of GenAI tools requires substantial organizational changes that address both technical and human factors.
Sauvola et al.~\cite{sauvola2024ase} analyzed the potential of GenAI for future software development and identified skill-gap challenges where developers lack necessary AI expertise, highlighting the need for investment in training programs to develop competencies in prompt engineering, AI output validation, and human-AI collaboration.
These organizational challenges have also led researchers to investigate GenAI adoption patterns.
Russo et al.~\cite{russo2024tosem} developed the Human-AI Collaboration and Adaptation Framework, a theoretical model designed to understand and predict GenAI tool adoption in software engineering. They found that compatibility factors---particularly, how well AI tools integrate within existing development workflows---serve as the primary driver of organizational adoption decisions. This finding challenges conventional technology acceptance theories~\cite{marangunic2015tam}, as traditional factors, such as perceived usefulness, social influence, and personal innovativeness, proved less influential than expected.

The integration of GenAI tools into complex software development workflows and ecosystems also involves legal considerations. 
Wintersgill et al.~\cite{DBLP:journals/pacmse/WintersgillSHCP24} examined OSS license compliance from the perspectives of legal practitioners, identifying challenges in managing compliance for traditional software components. 
As AI-generated code becomes more and more prevalent in open-source projects, OSS compliance frameworks may need to be adapted to address questions of attribution, licensing obligations, and intellectual property considerations for AI-generated content.

The limited analysis of current GenAI adoption policies represents an important research opportunity.
Our work contributes to filling this gap by examining how open-source projects are developing governance approaches to manage GenAI adoption and the specific risks and concerns (technical, ethical, and legal) that drive these policy decisions.

\subsection{GenAI Impact on Software Development}

A substantial amount of research has been conducted on quantifying the impacts of GenAI tools on software development processes and outcomes, moving beyond anecdotal evidence and developer perceptions.

Ziegler et al. conducted a large-scale empirical study examining GitHub Copilot’s effect on developer productivity~\cite{DBLP:journals/cacm/ZieglerKLRRSSA24}. Based on surveys and usage telemetry, they found that developers perceived productivity improvements when using Copilot, particularly for repetitive and routine coding activities, with the perceived magnitude of improvement varying considerably based on task complexity and context.

In 2024, GitClear analyzed over 150 million lines of code across GitHub repositories from 2020 to 2023 to assess the impact of AI-assisted development on code quality~\cite{gitclear2024}, attributing rising code churn to GenAI adoption and interpreting it as indicative of code that was incomplete or erroneous when initially committed.
The 2025 follow-up report~\cite{gitclear2025}, based on an expanded dataset of 211 million lines through 2024, reported a rise in code churn from 4.5\% in 2023 to 5.7\% in 2024, a 39.9\% drop in refactoring, and a 17.1\% increase in copy-pasted code.
The same report documented an eight-fold increase in duplicated code blocks during 2024 and reported that for the first time, copy-pasted lines exceeded moved lines within commits, indicating a fundamental shift away from code refactoring toward code duplication and raising concerns about growing technical debt and the long-term sustainability of AI-assisted coding.
However, our analysis of code churn in select GitHub repositories in which developers acknowledged GenAI usage reveals different patterns, suggesting that the relationship between AI assistance and code quality may be more nuanced than these industry reports indicate.

Pearce et al.~\cite{pearce2025cacm} conducted a security assessment of code contributions generated by GitHub Copilot across multiple programming languages and contexts.
They found systematic security weaknesses in AI-generated code, arguing that security issues introduced by GenAI tools stem from the models' training on publicly available code repositories, which inherently contain security flaws.
Asare et al.~\cite{asare2023ese} compared vulnerability rates between human-written and Copilot-generated code and found that, while the GenAI tool introduced security vulnerabilities, the rates were not higher than those introduced by human developers.
These findings suggest that security concerns with AI-generated code may reflect broader challenges in secure coding practices rather than AI-specific problems.

Our study adds to the existing body of knowledge by analyzing self-admitted GenAI usage across more than 200,000 OSS repositories and conducting a study of code churn, showing how OSS projects use GenAI tools and how their usage impacts development activity.

\section{Threats to Validity}

In this section, we discuss the threats to the construct, internal, and external validity of our study.

\subsection{Construct Validity}

Our reliance on self-admitted GenAI usage introduces two main threats.
First, we only captured the visible part of GenAI adoption in OSS projects.
Developers who use GenAI tools without leaving a trace remain outside of our analysis scope, meaning our findings represent a lower bound on actual GenAI adoption. 
Therefore, the observed patterns must be interpreted within this context, as they may not apply to all instances of GenAI-assisted software development.
Second, some self-admitted mentions introduce ambiguity in determining which portions of code were generated by GenAI tools.
When a developer comments that the code was ``generated by ChatGPT,'' this may refer to complete classes, functions/methods, code blocks, or merely an initial structure that was subsequently modified.
Although we always examined the whole context around a GenAI mention, we might have misclassified its scope and purpose in some instances.

The focus on self-admitted usage has implications for our answers to the research questions.
Developers may be more likely to acknowledge GenAI usage for mundane tasks such as translation rather than for core development tasks such as implementing complex business logic.
Therefore, the observed frequencies might reflect what developers feel comfortable disclosing rather than their actual usage (\textbf{RQ1}). 
Strict usage policies might lead to fewer self-admitted usages, as contributors might avoid disclosure (\textbf{RQ2}). 
Finally, unacknowledged usage before the first explicit mention could distort the pre-adoption baseline for our code churn analysis (\textbf{RQ3}).
However, at the same time, we consider self-admitted GenAI usage a useful analytical lens for studying the impact of GenAI adoption in open-source software projects.
Given our manual validation, the precision of the identified GenAI usages is high, even though the overall recall of all GenAI usages is low.

When calculating code churn, we used the first explicit mention of GenAI tools as a proxy for adoption timing, which may not accurately reflect when projects actually began using GenAI.
However, our generous analysis window of 360 days before and after this point and robustness checks (e.g., placebo tests) help accommodate potential discrepancies in adoption dates.
Although we focused on only one measure of code quality, the relevance of code churn as a metric was motivated by industry research. 
The GitClear 2025 report~\cite{gitclear2025} documented a rise in churn from 4.5\% in 2023 to 5.7\% in 2024, coinciding with the proliferation of GenAI-assisted development. This increase correlates with two related trends: a 39.9\% decline in ``moved'' code (indicating reduced refactoring) and a 17.1\% rise in ``copy/pasted'' code.
Previous research links reduced refactoring to higher defect rates~\cite{mohagheghi2004empirical} and code clones to increased technical debt~\cite{lerina2019investigating, feitosa2020code}.
Future work could expand our analysis by considering additional metrics.

\subsection{Internal Validity}

Our heuristic-based approach for detecting GenAI mentions may have produced false negatives, particularly for mentions using non-standard terminology or abbreviations.
We addressed this by developing comprehensive regular expressions, covering common naming variations, and conducting a thorough manual validation of the identified mentions.
We rely on manually annotated data, which may be miscoded due to the subjective nature of understanding the coding book.
To mitigate this threat and ensure consistency in our qualitative analysis, we implemented a rigorous manual review process with multiple raters in several rounds of independent coding, achieving high inter-rater reliability. 

The number of policies and guidelines we analyzed and the number of survey responses we received were relatively low.
However, even this limited data revealed diverse regulation approaches and opinions, motivating future research.

\subsection{External Validity}

We restricted our analysis to public repositories hosted on GitHub, focusing on five popular programming languages.
The self-admitted GenAI usage we studied might not reflect general GenAI usage---self-admitted or hidden---in other repositories, programming languages, or industrial software projects. 
However, the selected languages represent the most commonly used languages according to the 2024 GitHub Octoverse report~\cite{ghreport}. 
Furthermore, our filtering criteria for engineered software projects ensured that our findings reflect practices in actively maintained software projects. Moreover, our results may not generalize to other open-source platforms such as GitLab, which may have different norms and adoption patterns.

The developer survey that we conducted as part of \textbf{RQ2} received eight responses from contributors of one of the 12 projects that we identified to have explicit GenAI usage policies and guidelines. Their responses might not reflect the views of a broader developer population. Future work should extend this analysis, for example, by analyzing GitHub Discussion threads on AI regulation.
Finally, our focus on ChatGPT and GitHub Copilot might not capture the usage patterns of other GenAI tools that were released more recently, e.g., Claude Code.

\section{Conclusion}

This study introduced \emph{self-admitted GenAI usage}---explicit references to LLM-based tools such as ChatGPT and GitHub Copilot---as a novel lens for examining how generative AI is used in open-source software development.
In our mixed-methods study design, we first mined more than 200,000 GitHub repositories, isolating 1,292 true-positive GenAI mentions across 156 projects. Qualitative open coding of these instances and subsequent card sorting yielded taxonomies of 32 assisted tasks, 10 content types, and 11 usage purposes. 
We complemented this content analysis with a survey of project contributors and a systematic review of 13 project-level policies and guidelines.
In addition, we performed a regression-discontinuity (RDD) analysis of code churn in the 149 repositories that contained sufficient data to study the impact of GenAI adoption on open source software projects.
Based on our findings, we derive several actionable implications.

\textbf{Implications for software developers.} The \textbf{RQ1} results show that developers most often use GenAI for code generation, natural-language translation, and refactoring---with explicit acknowledgment as the dominant purpose. We also observed recurring follow-up actions, including bug fixes, refactorings, reversions, and deletions, triggered by earlier GenAI-generated content. Together with our \textbf{RQ3} task-level analysis, which shows that generation tasks are more rework-prone than optimization or maintenance, this suggests that GenAI output has a provisional nature. Developers should therefore plan explicit validation and revision steps after GenAI-assisted generation, particularly for code.
While acknowledgment of GenAI usage is common (\textbf{RQ1}), contextual metadata such as prompts, model versions, or scope of generation is rarely recorded. This gap points to a concrete opportunity for developers to adopt provenance conventions (e.g., structured comments or commit tags) that better support the corrective practices already observed in the data.
The \textbf{RQ2} findings show substantial variation in project-level GenAI governance. Developers contributing to different projects should therefore avoid assuming uniform norms and instead adapt their GenAI usage and disclosure practices to project-specific guidelines.

\textbf{Implications for project maintainers.} \textbf{RQ2} shows that maintainers are already actively regulating GenAI usage, but through heterogeneous approaches ranging from bans to selective encouragement. This diversity suggests that generic policies are unlikely to work. Instead, maintainers should align GenAI guidelines with project-specific factors such as contribution patterns, review capacity, and risk tolerance.
The \textbf{RQ3} results show no systematic increase in code churn after GenAI adoption, contradicting prominent industry claims. While some repositories exhibit increased churn, many show stable or decreasing trends, and effects vary widely across projects. This suggests that restrictive policies motivated solely by assumed quality degradation are not well supported by the evidence. Project-level monitoring, e.g., tracking churn or related indicators before and after GenAI adoption, offers a project-specific and evidence-based alternative.
The findings for \textbf{RQ1} highlight that GenAI is frequently used for PR descriptions and documentation, which are comparatively low-risk artifacts. Maintainers can act on this by explicitly encouraging GenAI usage in these areas to improve communication and review efficiency while limiting exposure to higher rework costs.

\textbf{Implications for tool builders and platform providers.} \textbf{RQ1} shows that developers already engage in voluntary disclosure of GenAI usage when the cost is low, indicating that the main limitation to transparency lies in the absence of supporting mechanisms rather than in developer opposition. Tool builders can act on this by offering built-in, optional disclosure mechanisms such as automatic annotations or commit templates that integrate with existing workflows.
\textbf{RQ3} indicates that GenAI-assisted code generation is more rework-prone than other uses. Tools could respond by explicitly supporting post-generation validation, for example, through prompts or workflow affordances that encourage human review of generated code.
At the platform level, support for project-specific GenAI policies, such as configurable disclosure requirements in pull requests, could help operationalize the heterogeneous governance approaches observed in \textbf{RQ2}.

\textbf{Implications for researchers.} Our curated dataset of 1,292 self-admitted GenAI mentions and the taxonomy derived in \textbf{RQ1} provide a foundation for scaling empirical studies via automated detection and large-scale mining. Researchers can build on this work by developing detectors for self-admitted GenAI usage and using them to study adoption and disclosure practices at scale.
\textbf{RQ3} highlights the importance of methodological granularity. Analyses that ignore project-specific adoption points or governance contexts risk drawing misleading conclusions. Future studies should therefore favor repository- and task-level designs when assessing the impact of GenAI usage, rather than relying on aggregate trends.
Because this study focuses on version-controlled artifacts, future research should extend the analysis to adjacent artifacts such as pull-request discussions and review comments. Such extensions would naturally build on our dataset, taxonomy, and churn analysis, helping to complete the picture of how self-admitted GenAI usage shapes collaborative software development.

\section*{ACKNOWLEDGMENTS}

We thank all survey participants for providing valuable insights for our research. The research contribution of Fabio Calefato was partially supported by the European Union, NextGenerationEU through the Italian Ministry of University and Research, Projects PRIN 2022 (``QualAI: Continuous Quality Improvement of AI-based Systems'', grant n. 2022B3BP5S, CUP: H53D23003510006). Tao Xiao is supported in part by JSPS Grant-in-Aid for JSPS Fellows 23KJ1589 and the Kayamori Foundation of Informational Science Advancement.

\balance
\bibliographystyle{IEEEtran}
\bibliography{main}

\end{document}